\documentclass[12pt]{article}
\usepackage{a4,epsfig,amssymb}
\usepackage{tabls}
\begin{document}

\begin{titlepage}

%%%%%%%%%%%%%%%%%%%%%%%%%%%%%%
\begin{flushright}
\begin{tabular}{l}
  LHCb-PHYS-2001-041 \\
  PHYSICS \\
  ADP-01-11/T446 \\
  PCCF-RI-01-04
\end{tabular}  
\end{flushright}
%%%%%%%%%%%%%%%%%%%%%%%%%%%%%%%
 
\null\vskip 0.5 true cm
%
%%%\vspace*{3cm}
\begin{center}
{{\huge \bf  $B^{0({\pm})}$ decays into two vector mesons  }} \\
\vskip 1.0 cm
{{\large \bf Physical motivations and a general method for simulations }}
\vskip 1 true cm
{ \Large Z.J.Ajaltouni$^1$\footnote{ajaltouni@in2p3.fr},
 O.Leitner$^{1,2}$\footnote{oleitner@physics.adelaide.edu.au},
 C.Rimbault$^1$\footnote{rimbault@clermont.in2p3.fr}}\\
\bigskip
{{\small \it $^1$ Laboratoire de Physique Corpusculaire de Clermont-Ferrand \\
IN2P3/CNRS Universit\'e Blaise Pascal \\
F-63177 Aubi\`ere Cedex France  \\
 $^2$ Department of Physics and Mathematical Physics and Special Research Centre for the Subatomic 
  Structure of Matter \\ 
  Adelaide University \\ 
  Adelaide 5005, Australia}}
\vskip 1 true cm
\end{center}
\vspace{1.5cm}

%%%%%%%%%%%%%%%%%%%%%%%%%%%%%%%%%%%%%%%%%%%%%%%%%%%%%%%%%

\begin{abstract}
In this paper, a complete description of the channels $B \rightarrow V_1 V_2$ is given. Emphasis is put
on the determination of the dynamical density matrix which elements are computed according to the Wilson
operator product expansions entering into the formulation of the weak effective hamiltonian. 
\vskip 0.3cm
Kinematical consequences related to the particular channel $B \rightarrow K^* {\rho}^0 (\omega)$ 
are described in details.
\end{abstract}

\end{titlepage}

\newpage

$~$

\newpage

%%%%%%%%%%%%%%%%%%%%%%%%%%%%%%%%%%%%%%%%%%%%%%%%%%%%%%%%%%%%%%%%%%%%%%%%%
%%\bigskip

\section{Introduction}

In a previous note \cite{ZJA}, an exhaustive study of the channel simulations:  

$$ B \to V_1 V_2, \ \ {\gamma} V, \ \  P V, \ \  P P,   $$

($V = 1^-, P = 0^-)$ has been performed by stressing the helicity formalism and its consequences.
General formulas have been established, notably those giving the final angular distributions in the case of
the production of two vector mesons decaying into pseudoscalar mesons.

 The squared modulus of the decay amplitude has the following form:
 
\begin{eqnarray}
 |A|^2 \propto  h_{\lambda,{\lambda}'} F_{\lambda,{\lambda}'}({\theta}_1) G_{\lambda,{\lambda}'} 
                       ({\theta}_2, \phi),
\end{eqnarray}

where (summation over $\lambda ({\lambda}')$ is omitted):

\begin{itemize}
 \item $h_{\lambda,{\lambda}'}$ is the matrix density element constructed from the weak effective hamiltonian 
  $H_w^{eff}$ taken between the initial state ($B_0$) and the final state $f$.
 \item  $F_{\lambda,{\lambda}'}({\theta}_1)$ and $G_{\lambda,{\lambda}'}({\theta}_2, \phi)$
 are the matrix elements related to the decays $V_1 \rightarrow a_1 + b_1$
 and $V_2  \rightarrow a_2 + b_2 $ respectively.
 \item ${\theta}_j$ is the polar angle of particle $a_j$ in the rest frame of
  the resonance $V_j$ while $\phi$ is the angular difference ${\phi}_2 - {\phi}_1$ , where ${\phi}_j$
  is the polar angle of $a_j$ in $V_j$ rest frame. 
\end{itemize}

$\lambda ({\lambda}')$ being the helicity state of the vector mesons; $ \lambda = -1, 0, +1$.
\vskip 0.5cm

 As it can be noticed, the essential parameters for the determination of the decay dynamics are the 
 {\it unknown matrix elements} $h_{\lambda,{\lambda}'}$;
 while the two other ones, $F_{\lambda,{\lambda}'}({\theta}_1)$
and  $G_{\lambda,{\lambda}'}({\theta}_2, \phi)$, are kinematic (or geometric) parameters because they are 
completely determined from the Wigner rotation matrices. The reader is referred to the note $99-051$ 
for a full kinematic description of the $B^0$ decay and the 
physical significance of the angles ${\theta}_{1,2}$ and $\phi$.

Before dealing with the mathematical determinations of the $h_{\lambda,{\lambda}'}$ elements, a simple
justification of the two vector meson channel is given below.

\section{Quantum numbers of the ${V_1}^0 {V_2}^0$ system}

 In the case of two vector meson $B^0$ decay, the most interesting case is the one related to 
 neutral mesons 
supplemented by the condition $ C|{{V^0}_i} \rangle = - |{{V^0}_i} \rangle $, where $C$ is the charge conjugation operator 
and $ {V^0}_i$ is a neutral vector meson {\it eigenstate} of $C$. Some examples of these channels are:

 $$ {\rho}^0  {\rho}^0, \ \ J/{\Psi}{\rho}^0, \ \  J/{\Psi} {\Phi}, \ \ {\Phi}{\Phi} \dots$$

These vector mesons have, in addition, the {\it parity} quantum number equal to $-1$. Noticing that the total 
angular momentum of the $V^0_1 V^0_2$ system: $ \vec J = {\vec \ell} + {\vec S}  = {\vec {s_B}}$ is equal
to zero and because the total
spin $ \vec S = {\vec {s_1}} + {\vec {s_2}} $,   with  $ s_1 = s_2 = 1$, 
 the orbital angular momentum can have three different values:  
$ \ell = S = 0, 1, 2 $.
\vskip 0.3cm
 
Thus, parity, charge conjugation and CP quantum numbers of the $ V^0_1 V^0_2 $ system can be computed:

$$ P(V^0_1 V^0_2) = {(-1)^2} {(-1)^{\ell}}, \ \ \   C(V^0_1 V^0_2) = (-1)^2, $$

   $$ \Downarrow  $$
   
 $$    CP(V^0_1 V^0_2) \ \ = \ \  {(-1)^{\ell}}. $$
   
\vskip 0.5cm   

 We are led to the important result that the CP value of $V^0_1 V^0_2$ is a {\bf mixing} of two different
eigenvalues $+1$ and $-1$  whatever the initial state ($ B^0$ or ${\bar {B^0}} $) is. A direct consequence of
this result is that {\it CP symmetry is not an exact one}. 

\vskip 0.3cm

The above relation does not hold for reactions involving a neutral $K^*$ like:

$$ B^0_d  \to  K^{*0} {\rho}^0, \ \ \  J/{\Psi} K^{*0}\dots $$

because $K^{*0}$ and ${\bar K}^{*0}$ are two {\it distinct particles}; 
$ C|K^{*0} \rangle = |\bar {K^{*0}} \rangle  \neq  |K^{*0} \rangle $. 

\par
However, it is worth noticing two interesting features for channels with an intermediate resonance like
$K^{*0} (\bar {K^{*0}})$:

$$ K^{*0} \to  K^+ {\pi}^-,   \ \ \ \  K^0 {\pi}^0,$$

$$ \bar {K^{*0}} \to  K^- {\pi}^+,   \ \ \ \  \bar {K^0} {\pi}^0. $$

 The decay channels are in the ratio ${2/3}$ and ${1/3}$ respectively.
On one hand, the sign of the charged kaon shows clearly the nature of the neutral $K^*$
 from which it comes and consequently 
the {\it flavour} of the original $B^0 (\bar {B^0})$. So, a neutral $K^*$ decay is a direct way
for {\it $B^0$ flavour tagging}.
\vskip 0.5cm

On the other hand, when a neutral $K^{*0} (\bar {K^{*0}})$ decays into $K^0 (\bar {K^0}) {\pi}^0$, 
the neutral kaon 
$K^0 (\bar {K^0})$ is not the true physical particle, because approximately $50\%$ of the 
$K^0 (\bar {K^0})$ go into $K^0_S$ and $50\%$ into $K^0_L$ respectively and the true {\it detectable} 
particle is $K^0_S$ which goes to ${\pi}^+ {\pi}^-$. 
\vskip 0.3cm
Thus, in the special channel:

 $$ B^0 ( \bar {B^0} )   \to  K^{*0} (\bar {K^{*0}}) {\rho}^0,   $$
 $$   \ \ \ \ \ \ \ \ \ \ \ \ \   \rightarrow   K^0_S {\pi}^0,                     $$

tagging the original $B^0$ is no longer possible but, the $K^0_S {\pi}^0$ being a {\it common final state}  
to both $B^0$ and $\bar {B^0}$,  
 the above relation $ CP = (-1)^{\ell}$ is still available \cite{RF}.  
\par
In the following, emphasis will be put on the channels
$K^{*0({\pm})}{\rho}^0{(\omega)}$ and the physical importance of the ${\rho}^0{(\omega)}$ mixing for the 
determination of {\bf CP} violation. 

%%%%%%%%%%%%%%%%%%%%%%%%%%%%%%%%%  ICI   %%%%%%%%%%%%%%%%%%%%%%%%%%%%%%%%%
%%%%%%%%%%%%%%%%%%%%%%%%% SCHEMAS TREE et PENGOUINS %%%%%%%%%%%%%%%%%%%%%%%%%%5
%%%%%%%%%%%%%            ===========================  %%%%%%%%%%%%%%%%%%%%%%%%%%
%%%
%%%
%%%%%%%%%%% FIGURE : DIAGRAMMES TREE-PENGOUIN K*0 RHO0  %%%%%%%%%%%%%%%%%%%%%%%%%%%
%%%
\begin{figure}[htbp]
  \begin{center}
 \mbox{\epsfig{file = 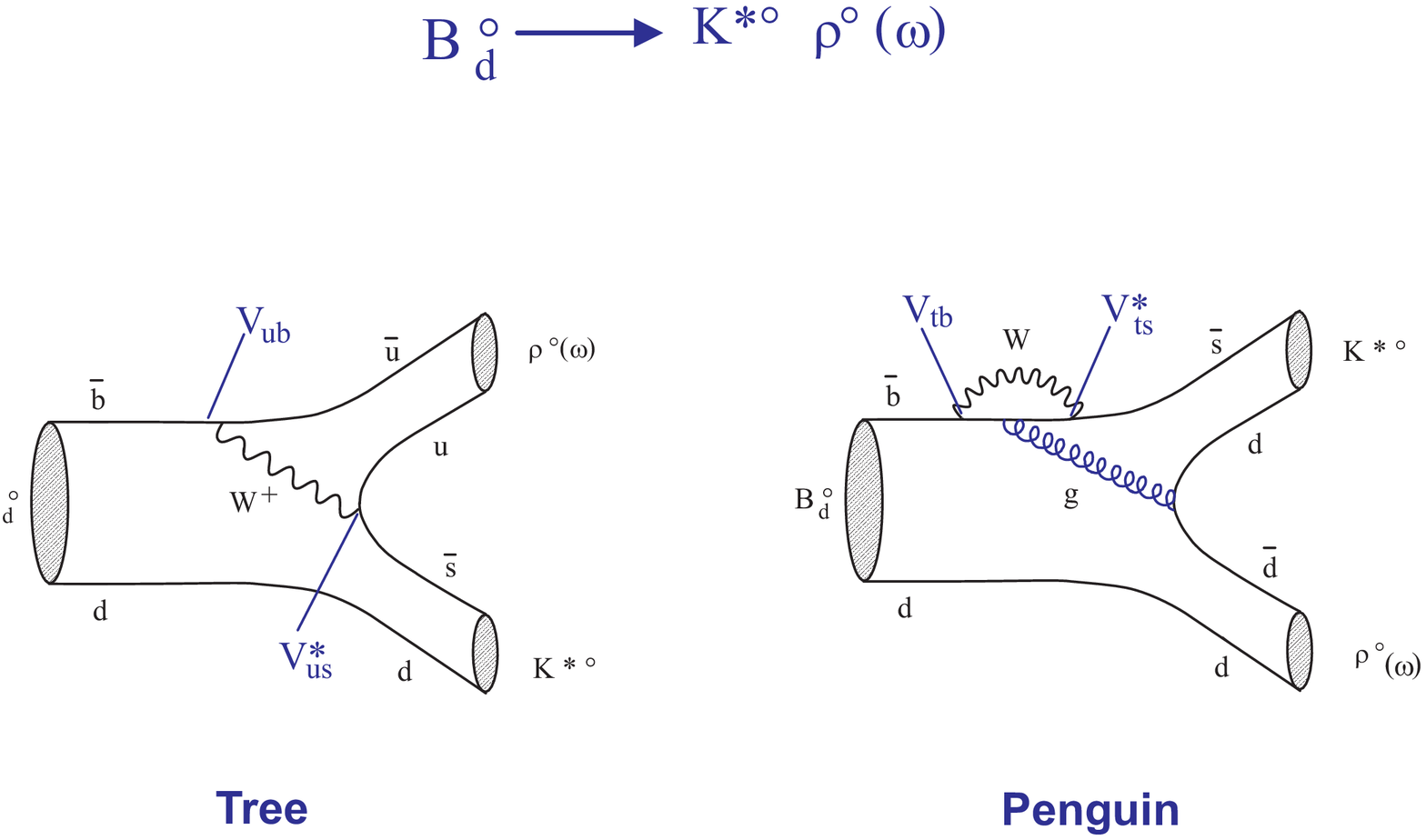, height = 10.0cm, width = 12.0cm}}
 \protect
 \caption{\it Tree and Penguin diagrams for the decay $B^0 \rightarrow K^{*0} {\rho^{0}(\omega)}$.}
  \end{center}
\end{figure}     

%%%%%%%%%%%%%%%%%%%%%%%%%%%%%%%%%%%%%%%%%%%%%%%%%%%%%%%%%%%%%%%%%%%%%%%%%%%%%%%%%%%%%%%%%%%%%%%%

\section{${\rho}^0{(\omega)}$ mixing and its consequence}

 It is well known from hadronic physics that the neutral isovector ${\rho}_8$ 
 and the isosinglet ${\omega}_8$ 
mix together, leading to the "true" physical resonances ${\rho}^0$ and $\omega$.
On the phenomenological level,  
this mixing is made possible because of the existence of a common final state to both ${\rho}^0$ and $\omega$  
decays \cite{PDG}:

  $$ {\rho}^0 \rightarrow  {\pi}^+  {\pi}^-,  \ \ \ \ \ (BR \approx  100\% ), $$
  $$ {\omega}  \rightarrow  {\pi}^+  {\pi}^-,  \ \ \ \ \ (BR \approx  2.2\% ). $$

In the same framework, it has been established that the $\pi \pi$ final state
interaction provides a phase shift $\delta$ which reaches $90^{\circ}$ when the $\pi \pi$ invariant mass 
is at the $\omega$ pole ($M_{\omega} = 782 \ MeV$) \cite{CONNELL}.    
\vskip 0.5cm

%%%%%%%%%%%%%%%%%%
 
%%%%%%%%%%% FIGURE : DIAGRAMMES TREE-PENGOUIN K*+  RHO0  %%%%%%%%%%%%%%%%%%%%%%%%%%%
%%%
\begin{figure}[htbp]
  \begin{center}
 \mbox{\epsfig{file = 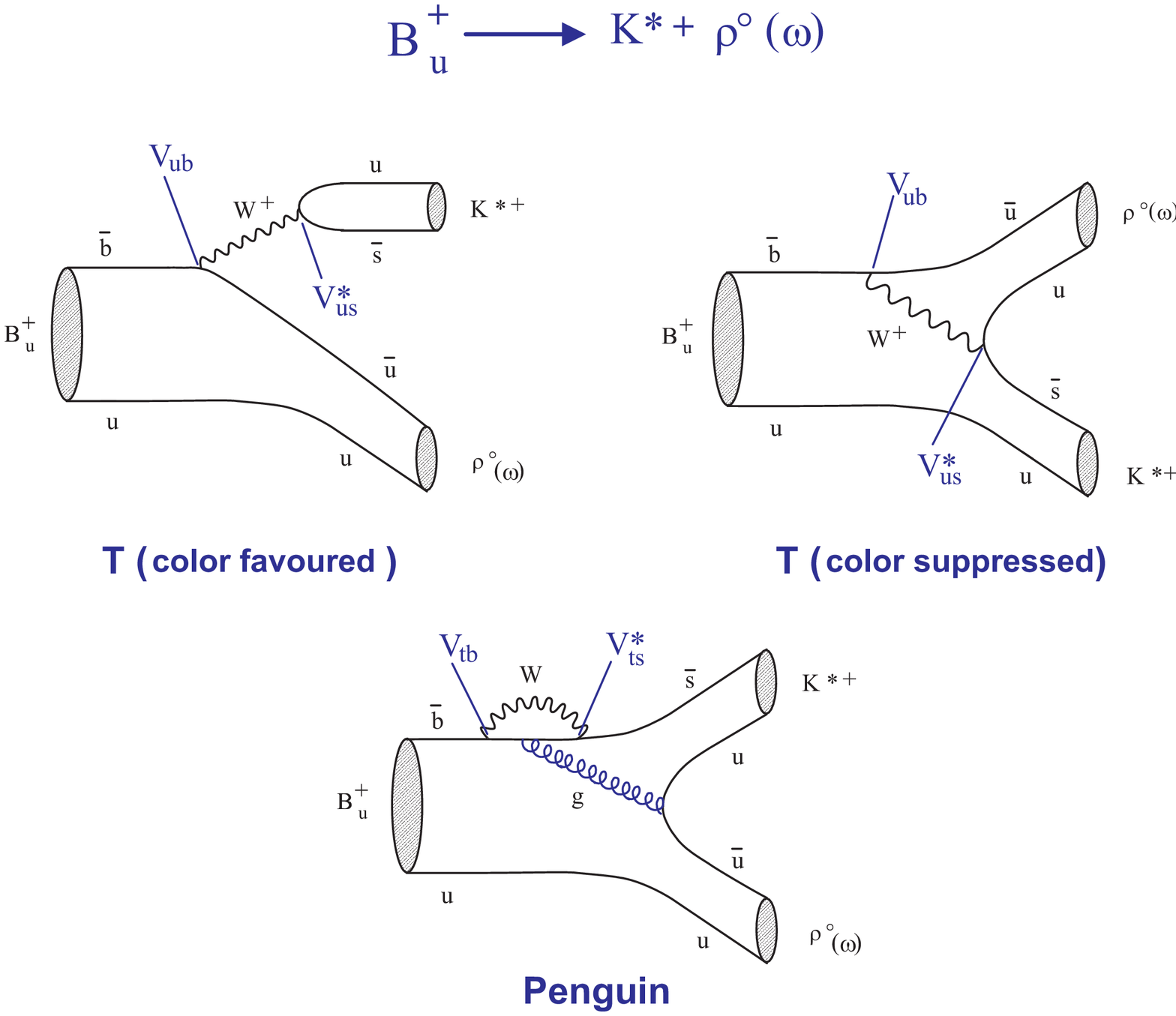, height = 14.0cm, width = 12.0cm}}
 \protect
 \caption{\it Tree and Penguin diagrams for the decay $B^+ \rightarrow K^{*+} {\rho^{0}(\omega)}$.}
  \end{center}
\end{figure}     

%%%%%%%%%%%%%%%%%%%%%%%%%%%%%%%%%%%%%%%%%%%%%%%%%%%%%%%%%%%%%%%%%%%%%%%%%%%%%%%%%%%%%%%%%%%%%%%%%%%%
%%%%%%%%%%%%%%%%%%%%%%%%%%%%%%%%%%%%%%%%%%%%%%%%%%%%%%%%%%%%%%%%%%%%%%%%%%%%%%%%%%%%%%%%%%%%%%%%%%%

This interesting physical property has important consequences in the case where a ${\rho^{0}}$ resonance is
produced in some $B^{0 {\pm}}$ decays like:

$$ B^0  \to   K^{*0}  {\rho}^0,  \ \ \ \  (Fig.1) $$

$$ B^+ \to   K^{*+} {\rho}^0,  \ \  B^- \to   K^{*-} {\rho}^0,  \ \ \ (Fig.2) $$

These decays require both tree (T) and penguin (P) diagrams. As it is emphasized in reference \cite{AWTHOMAS},
the amplitude $  A$ and $\bar  A$ respectively for $B^+$ and $B^-$ decays can be set in the 
following form:
\begin{eqnarray}
A   =  A^T + A^P  =  A^T{\left( 1 + r \exp{(i {\delta})} \exp{(i {\phi})} \right)},
\end{eqnarray}

\begin{eqnarray}
{\bar  A} = {\bar A}^T + {\bar A}^P = A^T{\left( 1 + r \exp{(i{\delta})} \exp{(-i{\phi})}\right)},
\end{eqnarray}

where:
\begin{eqnarray}
 r=\left|{\frac {A^P}{A^T}}\right|,
\end{eqnarray}

\begin{eqnarray}
{\bar A}^T = A^T, \ \ {\bar A}^P = {|A^P|} {\exp{(i{\delta})} \exp{(-i{\phi})}}. 
\end{eqnarray}

\vskip 0.3cm

Expressions of $A$ and $\bar A$ displayed above suppose that final state interactions (FSI) arise essentially
from the penguin diagrams; this hypothesis is supported by the fact that, to order $G_F {\alpha}_s$ 
($G_F$ and ${\alpha}_s$ are respectively the Fermi constant and the QCD fine-structure constant), 
the {\it absorptive} part of the transition amplitude is obtained from the penguin diagrams 
\cite{BANDERLYPKIN}.
\par
In the special case of ${\rho}^0 - \omega$ mixing, another hypothesis is made using more intuitive arguments:
the phase shift due to the mixing is included in the FSI and it is {\it predominating} at the $\omega$ pole,
justifying the above expressions of $A$ and $\bar A$ that the phase shift $\delta$ is principally the one
generated by the ${\rho}^0 - \omega$ mixing.     

By $CP$ transformation, the strong phase $\delta$ remains unchanged while the weak phase $\phi$, 
which is related to the CKM matrix elements, changes sign.
 Thus, the asymmetry parameter $a_{CP}^{dir}$ 
which can reveal {\it direct CP violation} can be deduced in the following way:

\begin{eqnarray}
a_{CP}^{dir} = {\frac {A^2 -{\bar A}^2}{A^2 +{\bar A}^2}} = \frac{-2 \ {\sin {\delta}} \ {\sin {\phi}}}{1 + r^2
+2r \ {\cos {\delta}} \ {\cos {\phi}}}. 
\end{eqnarray}

It is straightforward to notice that the parameter $a_{CP}^{dir}$ depends both on the strong phase 
{\it and} the weak phase and, consequently, 
 the maximum value of $a_{CP}^{dir}$ can be reached if $\sin {\delta} = 1$, 
which allows us to state that
the strong final state interaction (FSI) among pions coming from the ${\rho}^0 - \omega$ decays
 {\it enhances} the
direct $CP$ violation in the vicinity of the resonance $\omega$ mass.

%%%%%%%%%%%%%%%%%%%%%%%%%%%%%%%%%%%% mixing RHO-OMEGA %%%%%%%%%%%%%%%%%%%%%%%%%%%%%%%%%%%%%%
%%%
\begin{figure}[htbp]
  \begin{center}
 \mbox{\epsfig{file = 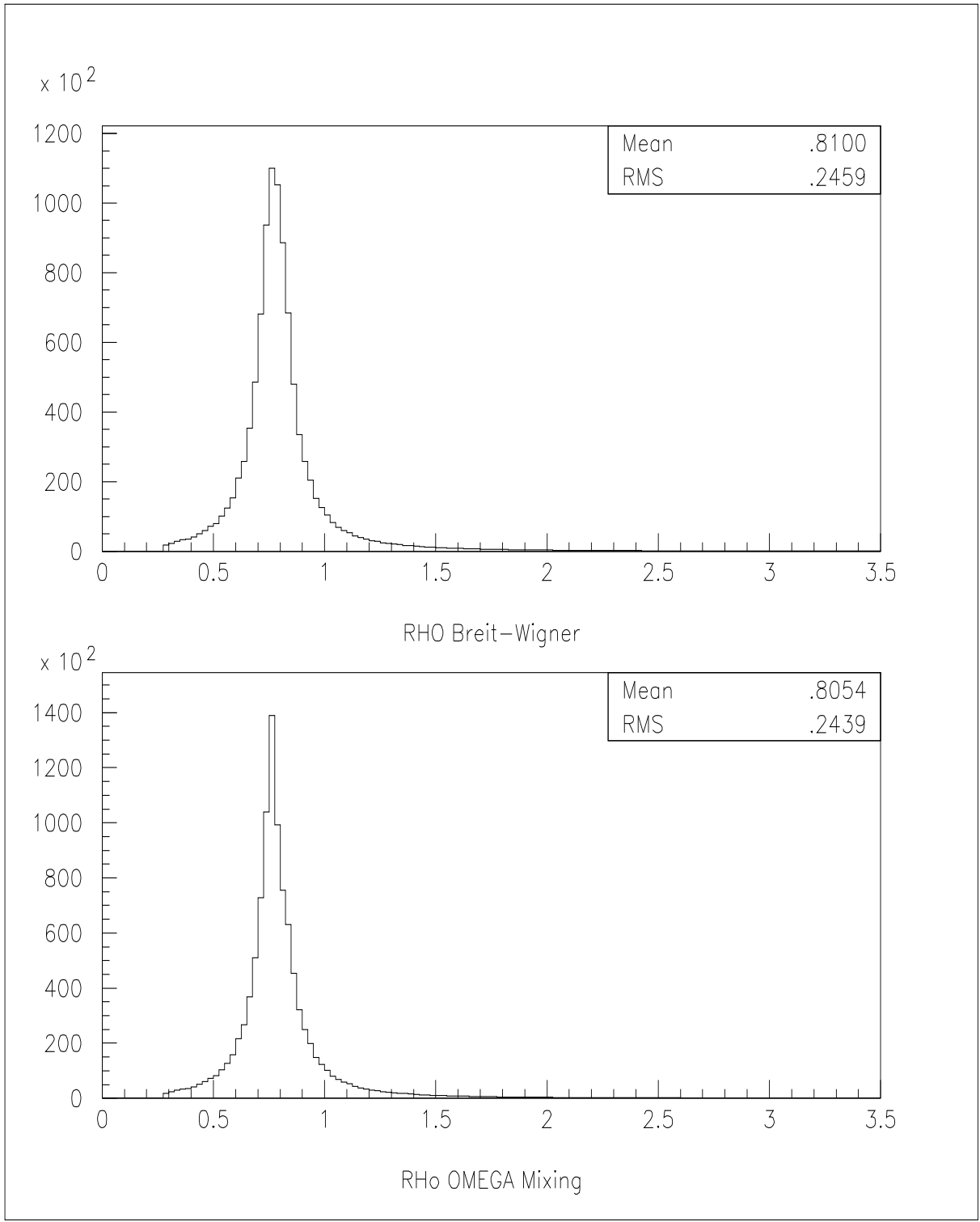, height = 20.0cm, width = 16.0cm}}
 \protect
\caption{\it Spectrum (in GeV/$c^2$) of ${\rho}^0$ Breit-Wigner (upper histogram) and ${\rho^{0}}- {\omega}$ mixing (lower
histogram).}
  \end{center} 
\end{figure}     

%%%%%%%%%%%%%%%%%%%%%%%%%%%%%%%%%%%%%%%%%%%%%%%%%%%%%%%%%%%%%%%%%%%%%%%%%%%%%%%%%%%%%%%%%%%%%%%%%%%%

\subsubsection*{Simulation of the ${\rho}^0 - \omega$ mixing}

 A simple and phenomenological relation describing the amplitude of the  
 ${\rho}^0 - \omega$ mixing  is used for the Monte-carlo simulations \cite{LANGACKER}. In the 
 ${\rho}^0$ Breit-Wigner, the (${\rho}^0$) propagator is replaced by the following one:

 \begin{eqnarray} 
 {\cal A} = {\frac{1}{s_{\rho}}}+{\frac{T_{\omega}}{T_ {\rho}}} {\frac{{\Pi}_{\rho \omega}}{{s_{\rho}}{s_{\omega}}}},
\end{eqnarray}
where

\begin{itemize}
 \item $1/s_V = 1/(s - {M_V}^2 + i {{\Gamma}_V}{M_V}) $ is the $V$ resonance propagator, $M_V$ and ${\Gamma}_V$
 being respectively the mass and the width of the resonance $V$.
 \item $T_{\omega}$ and $T_{\rho}$ are respectively the $\omega$ and $\rho$ production amplitudes.
 \item ${\Pi}_{\rho \omega}$ is the mixing parameter for which recent values come from $e^+ e^-$ annihilations:
\end{itemize}

 $ \Re e {({\Pi}_{\rho \omega})} = -3500 \pm 300  \ \ MeV^2 $  and $ \Im m{({\Pi}_{\rho \omega})} = -300  \pm 300  \ \ MeV^2.$
 
\vskip 0.5cm

Due to the same physical processes which enter into the production of the ${\rho}^0$ and $\omega$ resonances 
(they are both made out from $u {\bar u}$ and $d {\bar d}$ quark pairs with the same weight $1/2$), it seems
natural to choose ${T_{\omega}}/{T_ {\rho}} = 1 $. So, the squared mass distribution of the $\pi \pi$ system
becomes simplified and it is given by:

\begin{equation}
{d\sigma}/{dm^2}  \propto   {|{\cal A}({{\rho^{0}} ({\omega})})|}^2, 
\end{equation} 
 
 where ${\cal A}$ is the amplitude of the two Breit-Wigner given above and $m$ is the $\pi \pi$ invariant mass.

In Figure $3$,  are displayed the  $\pi \pi$ invariant mass spectra for the $\rho^{0}$
Breit-Wigner and the ${\rho}^0 - {\omega}$ mixing respectively. Because of the very narrow $\omega$ width
(${\Gamma}_{\omega} = 8.44 \ MeV $), we notice a high and narrow peak at the $\omega$ pole 
($\approx 782 \ MeV $).

%%%%%%%%%%%%%%%%%%%%%%%%%%%%%%%%%%%%%%%%%%%%%%%%%%%%%%%%%%%%%%%%%%%%%%%%%%%%%%%%%%%%%%%%%%%%%%
\section{Dynamics of the $B \rightarrow V_1 V_2 $ decay}

 The formalism describing the $B^{0 {(\pm)}}$ decay into two vector mesons is derived from the
general formalism  
related to the hadronic weak decay of a heavy meson (or heavy quark). 
It is based on the new concepts introduced
by the Heavy Quark Effective Theory ({\bf HQET}) which involves additionnal symmetry due to the high mass of
the heavy quark ($b$ or $c$ quark) \cite{NEUBERT}.
 Technical calculations require a weak effective hamiltonian,
$H_w^{eff}$, by using the "Operator Product Expansions" ({\bf OPE}) pioneered by Wilson and which involve
field operators describing both {\it tree} and {\it penguin} diagrams, the last ones include both QCD and
electroweak penguins (Figures 1 and 2).

The general form of  $H_w^{eff}$ is given by:

%%% $$ H_w^{eff} = {\frac {G_F}{\sqrt 2}} {\sum}_{n = 1,10} {C_n({\mu})} {O_n({\mu})}   $$

\begin{equation}
{H_w}^{eff} = {\frac {G_F}{\sqrt 2}} {{\sum}_{q=d,s} \Big(V_{ub}V_{uq}^*{(c_1
    O_1 +c_2 O_2)}
-V_{tb}V_{tq}^* {{\sum}_{i=3}^{10} {c_iO_i}} \Big)},
\end{equation} 
 
%% where ${c_n({\mu})}$ are the Wilson coefficients and ${O_n({\mu})}$ 

where $c_i$ are the Wilson coefficients and $O_i$ are field operators with dimension
  $d \geq 4$; they are computed at an energy scale $\mu$ which is identified, here, with the $b$ quark mass 
  $m_b$.

In the case of charmless $B$ decays, Wilson coefficients have been calculated by Buchalla et al 
\cite{BUCHALLA}.
These coefficients represent the {\it perturbative part} of the weak hamiltonian, 
they are estimated by the Renormalization Group techniques 
and their values depend on the renormalization scheme which is used. Their physical significance 
 is the
{\it weight} of each field operator $O_i{(\mu)}$ entering in the weak hamiltonian $H_w^{eff}$. 
From reference \cite{DESHPANDE}, the values of $c_i$ which have been computed at the energy scale 
$\mu = m_b $ are:

\vspace{-1.0em}
\begin{eqnarray*}
\begin{array}{ll}
\vspace{0.5em}
 c_{1}=-0.3125,\;\;\;\; c_{2}=1.1502,   \\
\vspace{0.5em}
 c_{3}=0.0174 ,\;\;\;\; c_{4}=-0.0373,   \\
\vspace{0.5em}
 c_{5}=0.0104 ,\;\;\;\; c_{6}=-0.0459,  \\
\vspace{0.5em}
 c_{7}=-1.050 \times 10^{-5}, \;\;\;\; c_{8}=3.839 \times 10^{-4},  
\end{array}
\end{eqnarray*}
\vspace{-2.4em}
\begin{eqnarray}
\!\!\!\!\!\!\!\! \!\!\!\!\!\!\!c_{9}=-0.0101, \;\;\;\; 
c_{10}=1.959 \times 10^{-3}.  & &  
\end{eqnarray}
%

%$$ c_1 = -0.3125, \ \ c_2 = 1.1502,  $$
%$$ c_3 = 0.0174, \ \ c_4 = -0.0373, \ \ c_5 = 0.0104, \ \  c_6 = -0.0459, $$
%$$ c_7 = -1.050 \times 10^{-5}, \ \ c_8 = 3.839, \times 10^{-4} $$ 
%\begin{equation}
%c_9 = -0.0101, \ \  c_{10} = 1.959 \times  10^{-3}.          \\
%\end{equation}

The first two coefficients, $c_1$ and $c_2$, are related to the {\it tree} diagrams and they show clearly 
their dominance with respect to the {\it penguin} ones. Coefficients $c_3 - c_6$ correspond to QCD penguin
operators while $ c_7 - c_{10}$ are related to the EW ones.
\par
However, those values of $c_i$ must be modified when renormalization of operator $O_i$ at one-loop 
order is taken into account.

\vskip 0.3cm

Detailed expressions of operators $O_i{(\mu)}$ and their physical interpretation are given 
 in reference \cite{IJMP}. 
 
\vskip 0.3cm

Thus, a general form for the weak decay amplitude into a final state $f$ can be expressed like:

\begin{equation}
 A(B^0 \rightarrow f) = \langle f|H_w^{eff}|B^0 \rangle = {\frac {G_F}{\sqrt 2}} {\sum_{i=1}^{10}}{\sum_{q= d,s}}
{{\lambda}_q}^i {c_i({\mu})}{\langle f|O_i({\mu})|B^0 \rangle},
\end{equation}

where ${{\lambda}_q}^i$ is the product of two CKM matrix elements: $V_{ub}V_{uq}^*$ (for $i=1,2$) or
 ${(V_{tb}V_{tq}^*)}$ (for $i=3 ,\dots, 10$).
%%% with $q = d,s$ and $i = 1-10$,
%%%{V_{CKM}^i}{c_i({\mu})}{<f|O_i({\mu})|B^0>} $$
%%%$V_{CKM}^i$ being the CKM matrix element.

\vskip 0.4cm

The hadronic matrix elements $ \langle f|O_i({\mu})|B^0 \rangle$ represent the {\it non-perturbative contribution}
 to the amplitude $A(B^0 \rightarrow f)$.
 Usually, they are estimated according to some specific models: 
Non Relativistic Quark Model (NRQM), Form Factor models (BSW) and especially the Lattice QCD calculations.

In the following, calculation of the hadronic matrix elements is performed in the framework of the BSW 
model \cite{BSW} from which form factors are derived by the knowledge of the hadronic wave functions for both
initial and final states.

\section{Determination of the density-matrix elements}

The $B^0$ decay into two vector mesons requires the helicity formalism which has been intensively used in the 
previous paper \cite{ZJA}. To each vector meson (spin $1$) is assigned a set of three
polarization 4-vectors defined in this way:

\begin{equation}
{\epsilon}_1 = (0, \vec {{\epsilon}_1}), \ \ {\epsilon}_2 = (0, \vec {{\epsilon}_2}), \ \  {\epsilon}_3
={\left(|{\vec k}|/m , E {\hat {k}}/m \right)},
\end{equation}

and verifying the following relations:
\begin{equation}
 {{\epsilon}_i}^2 = -1, \ \ {\epsilon}_i \cdot {\epsilon}_j = 0, \ \  \mathrm{with} \  i \neq j,
\end{equation}

where $m, E, \vec k$ are respectively the mass, the energy and the momentum of the vector meson; $\hat {k}$ is
defined as the unit vector along the vector momentum, $\hat {k} = {\vec k}/{|\vec k|}$.
\vskip 0.5 cm
The three vectors $\vec {{\epsilon}_1}, \vec {{\epsilon}_2}$ and $\vec {{\epsilon}_3} = {E {\hat {k}}}/m$ 
form an orthogonal basis; 
${\epsilon}_1$ and ${\epsilon}_2$ are called the {\it transverse polarization} vectors while  
$\vec {{\epsilon}_3}$ is the {\it longitudinal polarization} one. 
\par

From that basis, an {\it helicity basis} is defined according to:

\begin{equation}
{\epsilon}(+) = \frac{\left({\epsilon}_1 + i {\epsilon}_2 \right)}{\sqrt 2}, \ \ {\epsilon}(-) =
\frac{\left({\epsilon}_1 - i {\epsilon}_2 \right)}{\sqrt 2}, \ \   {\epsilon}(0) = {\epsilon}_3.
\end{equation}

 These 4-vectors are {\it eigenvectors} of the helicity operator $\mathcal {H}$ with the eigenvalues 
$ \lambda = +1, -1$ and $0$ respectively. For a clear account of the helicity basis for a spin $1$ particle,
 the reader can consult the book of Dewitt-Smith \cite{DEWITT}.  
\vskip 0.5 cm

In the case of two vector mesons coming from the $B$ decay, their 4-momenta are defined in the $B$ rest frame
and their corresponding polarization vectors are {\it correlated} because $\hat {k_1} = - \hat {k_2}$.
For an explicit calculation of their spatial components, see the appendix A.  
\vskip 0.5cm

The weak hadronic amplitude is then decomposed on the helicity basis according to the general formalism
developed by the authors BSW \cite{BSW}. This method allows one to obtain two interesting results: 
\par

$\bullet$ the contribution of the {\it tree} and {\it penguin} operators to the global amplitude via the
  helicity states.  
\par
  
$\bullet$ the total contribution of each helicity state. \\

A way of illustrating this method is to study the channel: 
$B^0{(\bar B^0)} \rightarrow K^{*0} {(\bar K^{*0})}  {\rho}^0$. 
     
\vskip 0.3cm 
%%%Full calculations have been performed by one of us (O.L.) \cite{OLIVIER} by using the 
%%%analytical method outlined in reference \cite{AWTHOMAS} ~:

$(i)$ First of all, the mass of each resonance
($K^{*0}$ and ${\rho}^0$ ) is generated according to a relativistic Breit-Wigner: 

%%%%%%%%%%%%%%%     FORMULE DE LA BREIT-WIGNER   %%%%%%%%%%%%%%%%
\begin{equation}
\frac{d\sigma}{dM^2} \ \ = \ \  C \frac{\Gamma_R M_R}{{(M^2-{M^2_R})}^2 + {(\Gamma_R M_R)}^2}, 
\end{equation}

$C$ being a normalization constant.

\vskip 0.6cm
 
$(ii)$ The weak hadronic matrix element is expressed as the {\bf sum} of three helicity matrix elements;  
each one of the form, 
$ H_{\lambda} = \langle V_1 V_2 | {H_w}^{eff} | B \rangle $, is defined by gathering all the Wilson coefficients of
both tree and penguin operators. Linear combinations of those coefficients arise like:
 $c_{t1}^{\rho}, \  c_{p1}^{\rho}, \ $ and  $c_{p2}^{\rho}$ (see Appendix B) 
and the {\it helicity amplitude} $H_ {\lambda}$ gets the following expression:

\vskip 0.3cm

$$H_{\lambda} =\Big(V_{ub}V_{us}^{*}c_{t_{1}}^{\rho}-V_{tb}V_{ts}^{*}c_{p_{2}}^{\rho}\Big)\bigg\lbrace \beta_{1}\varepsilon_{\alpha \beta \gamma \delta}\epsilon_{K}^{*\alpha}(\lambda)\epsilon_{\rho}^{*\beta}(\lambda)P_{B}^{\gamma}P_{K}^{\delta}$$
$$+i\Big(\beta_{2}\epsilon_{K}^{*}(\lambda)\epsilon_{\rho}^{*}(\lambda) - \beta_{3}(\epsilon_{K}^{*}(\lambda).P_{B})(\epsilon_{\rho}^{*}(\lambda).P_{B})\Big)\bigg\rbrace$$
$$ + \Big(-V_{tb}V_{ts}^{*}c_{p_{1}}^{\rho}\Big)\bigg\lbrace \beta_{4}\varepsilon_{\alpha \beta \gamma \delta}\epsilon_{\rho}^{*\alpha}(\lambda)\epsilon_{K}^{*\beta}(\lambda)P_{B}^{\gamma}P_{\rho}^{\delta},$$
\begin{equation}
+i\Big(\beta_{5}\epsilon_{\rho}^{*}(\lambda)\epsilon_{K}^{*}(\lambda) - \beta_{6}(\epsilon_{\rho}^{*}(\lambda).P_{B})(\epsilon_{K}^{*}(\lambda).P_{B})\Big)\bigg\rbrace
\end{equation}

\vskip 0.5cm

with: 

\ $\bullet$\ $\varepsilon_{\alpha \beta \gamma \delta}$: antisymmetric tensor in the Minkowski space. \\
 
\ $\bullet$\ $\beta_{1,4}=\frac{G_{F}}{2} f_{\rho,{K}}m_{\rho,{K^*}}\frac{2}{m_{B}+m_{K^*,{\rho}}}V^{B\to K^*,{\rho}}(m^{2}_{\rho,{K^*}})$.  \\  

\ $\bullet$\ $\beta_{2,5}=\frac{G_{F}}{2} f_{\rho, K}m_{\rho, K^*}(m_{B}+m_{K^*,{\rho}})A_{1}^{B\to K^*,{\rho}}(m^{2}_{\rho, K^*})$.    \\

\ $\bullet$\ $\beta_{3,6}=\frac{G_{F}}{2} f_{\rho, K}m_{\rho, K^*}\frac{2}{m_{B}+m_{K^*,{\rho}}}A_{2}^{B\to K^*,{\rho}}(m^{2}_{\rho, K^*})$.  \\

\ $\bullet$ $f_{K}$, $f_{\rho}$: respectively $K^{*0}$ and $\rho^{0}$ decay constants. \\

\ $\bullet$ $V^{B\to K^*,{\rho}}$, $A_{i}^{B \to K^*,\rho}$: respectively Vector and Axial form factors (see Appendix C). \\

\ $\bullet$ $\epsilon_{K, \rho}(\lambda)$: $K^{*0}$,  $\rho^{0}$ polarization vectors expressed in the $B$ rest frame.

\vskip 0.5cm

It is worth noticing that the tensorial terms which enter $H_{\lambda}$ become simplified in the $B$
rest frame because the $B$ 4-momentum is given by $P_B = {(m_b, {\vec 0})}$. Then, using the
orthogonality properties of ${\epsilon}_j{(\lambda)}$, the helicity amplitude $H_{\lambda}$ acquires a 
much simpler expression than above:

\begin{equation}
H(\lambda) = iB(\lambda)(V_{ub}V_{us}^{*}c_{t_1}^{\rho}-V_{tb}V_{ts}^{*}c_{p_2}^{\rho})+
iC(\lambda)(-V_{tb}V_{ts}^{*}c_{p_1}^{\rho}),
\end{equation}

\vskip 0.6cm

 with:
 
$$B(0)=\beta_{2}\frac{m_{B}^2 -(m_{K}^2 + m_{\rho}^2 )}{2m_{K}m_{\rho}} - \beta_{3}\frac{|\vec{p}|^{2}m_{B}^2 }{m_{K}m_{\rho}},$$

$$C(0)=\beta_{5}\frac{m_{B}^2 -(m_{K}^2 + m_{\rho}^2 )}{2m_{K}m_{\rho}} - \beta_{6}\frac{|\vec{p}|^{2}m_{B}^2}{m_{K}m_{\rho}},$$

$$ B(\pm{1}) = \mp \beta_{1}m_{B}|\vec{p}| - \beta_{2},$$

\begin{equation}
C(\pm{1}) = \mp \beta_{4}m_{B}|\vec{p}| - \beta_{5},
\end{equation}

$|{\vec{p}}|$ being the common momentum to $V_1$ and $V_2$ particles in the $B$ rest frame.

\vskip 0.6 cm
 
$(iii)$ Expressing the CKM matrix elements according to Wolfenstein parametrization \cite{EPJC15}:
 
\begin{equation}
V_{CKM} = \left[
                 \begin{array}{ccc}
                 1-\frac{\lambda^2}{2} \ & \ \lambda \ & \
A\lambda^3(\rho-i\eta) \\
                 -\lambda \ & \ 1-\frac{\lambda^2}{2}\  & \ A\lambda^2
\\
                 A\lambda^3(1-\rho-i\eta)\  & \  -A\lambda^2 \ & \  1 \\

                \end{array}
          \right]    \ \ \ \ \  + \ \ O{({\lambda}^4),}        
\end{equation}

\vskip 0.6cm

where we use\cite{ACHILLE}:

{  \center
\ $A = 0.815$, \ \ \ \ $\lambda = 0.2205$: well known\\

\ {\bf $0.09<\rho<0.254$, \ \ \ \ $0.323<\eta<0.442$.}\\ }

\vskip 0.6cm 

 Taking into account the 
preceding relations, we arrive at the final form for the amplitudes $H_{\lambda}$:

{
$$H {0 \choose \pm{1}} = A\lambda^{2} \Bigg \lbrace \bigg \lbrack \Big(\eta \lambda^2 c_{t_1}^{\rho} - \Im{m(c_{p_2}^{\rho})}\Big)B {0 \choose \pm{1}} - \Im{m(c_{p_1}^{\rho})}C{0 \choose \pm{1}}\bigg \rbrack$$
\begin{equation}
+ \ i\ \bigg \lbrack \Big( \rho \lambda^{2}c_{t_1}^{\rho} + \Re{e(c_{p_2}^{\rho})}\Big)B{0 \choose\pm{1}} + \Re{e(c_{p_1}^{\rho})}C{0 \choose \pm{1}} \bigg \rbrack \Bigg \rbrace,
\end{equation}
}

 from which the density-matrix elements $h_{{\lambda}, {\lambda}'}$ can be derived automatically;
 
   $\Rightarrow  h_{{\lambda}, {\lambda}'} = H_{\lambda} {H_{{\lambda}'}}^* $.
   
\vskip 0.4cm

Due to the {\it hermiticity} of the matrix ($h_{{\lambda}, {\lambda}'}$), only six elements must be calculated
and, furthermore, a normalization condition is applied:
\begin{equation}
 N  \ {\left(h_{++} + h_{00} + h_{--}\right)}  \ \  = \ \  1,
\end{equation}

($N$ being the normalization constant) which makes easier the comparison 
of the modulus of the different matrix elements.

\vskip 0.6 cm

In the next histograms (Fig.4 - Fig.9) are displayed the spectra 
of $h_{{\lambda}, {\lambda}'}$ for different values 
of the Wolfenstein parameters $\rho$ and $\eta$. In our study, these spectra are obtained for the four couples
of values: $(0.09, 0.323)$; $(0.09, 0.442)$; $(0.254, 0.323)$  and $(0.254, 0.442)$.
But, due to the fact that some density matrix elements do not
vary too much with $\rho$ and $\eta$, in most cases only the spectra corresponding to the first couple 
of values are shown. All the histograms correspond to a sample of $20000$ generated events.    
 
\vskip 0.4cm
 It is important to notice that large spectrum of values for $h_{{\lambda}, {\lambda}'}$ are obtained 
and {\it not single ones} because of the broad range of both the ${\rho}^0$ resonance mass and the
common momentum $|{\vec p}|$  (see the analytical expressions of $B({\lambda})$ and
$C({\lambda})$ given above). 

\begin{itemize}
 \item Whatever the values of $\rho$ and $\eta$ are, the dominant value of $h_{++} = {|H_{+1}|}^2$
  is $\leq 10^{-2}$, numerical result which is proved too by complete analytical calculations. Thus  
 the dominant polarization state is the {\bf longitudinal} one because 
  $h_{00} = {|H_{0}|}^2 \ \  \geq  60\%$, its mean value being around $85\%$ (Fig.4).
 \item Due to the tiny value of $|H_{+1}|$, the modulus of the non-diagonal elements $h_{+-} = {H_+}{H_-}^*$
 and $h_{+0} = {H_+}{H_0}^*$ are usually smaller than $0.2$; while the modulus of $h_{-0} = {H_-}{H_0}^*$
 can reach $0.5$ (Fig.5).
\end{itemize}

Fig.6 and Fig.7 display the variations of the diagonal matrix elements $h_{--}, h_{00}$ and $h_{++}$ with respect
to the four sets of $\rho$ and $\eta$ values: it can be seen that $h_{++}$ has always a tiny value and
$ h_{00}$ is always dominant. Other physical features appear: $h_{00}$ is very {\it sensitive}  
to the parameter $\eta$; its spectrum is rather wide
 for ${\eta} = 0.323$, while it is bounded between $0.8$ and $1.0$ for ${\eta} = 0.442$. 
 For a fixed value of $\eta$, no noticeable variation with the parameter $\rho$ is seen. 
\vskip 0.3 cm
 Fig.8 shows the real and imaginary parts of the non-diagonal elements $h_{-0}, h_{+-}, h_{+0}$ 
respectively for $\rho = 0.09$ and $  \eta = 0.323$. It is worth noticing that both real and
imaginary parts of $h_{+-}$ and $h_{+0}$ are too small and close to zero.
\vskip 0.3 cm
Due to the importance of $h_{+-}$ matrix element in the $\phi$ angle distribution (see Section 6),
a full study of both real and imaginary parts of $h_{+-}$ has been done. Fig.9 shows the corresponding
spectra according to the values of $\rho$ and $\eta$. It can be deduced that the real and imaginary parts have
very similar distributions and both are dominated by small values ($\leq 0.05$).

%%%\newpage

%%%%%%%%%%%%%%%%%%%%%%%%%%%  HISTOGRAMMES des Valeurs DIAGONALES%%%%%%%%%%%%%%%%%%%%%%%%%%

%%%%%%%%%%%%%%%%%%%%%%%%%%%%%%%%%%%%%%%%%%%%%%%%%%%%%%%%%%%%%%%%%%%%%%%%%%%%%%%%%%%%%%%%%%%%%%%%
%%%
\begin{figure}[htbp]
  \begin{center}
 \mbox{\epsfig{file = 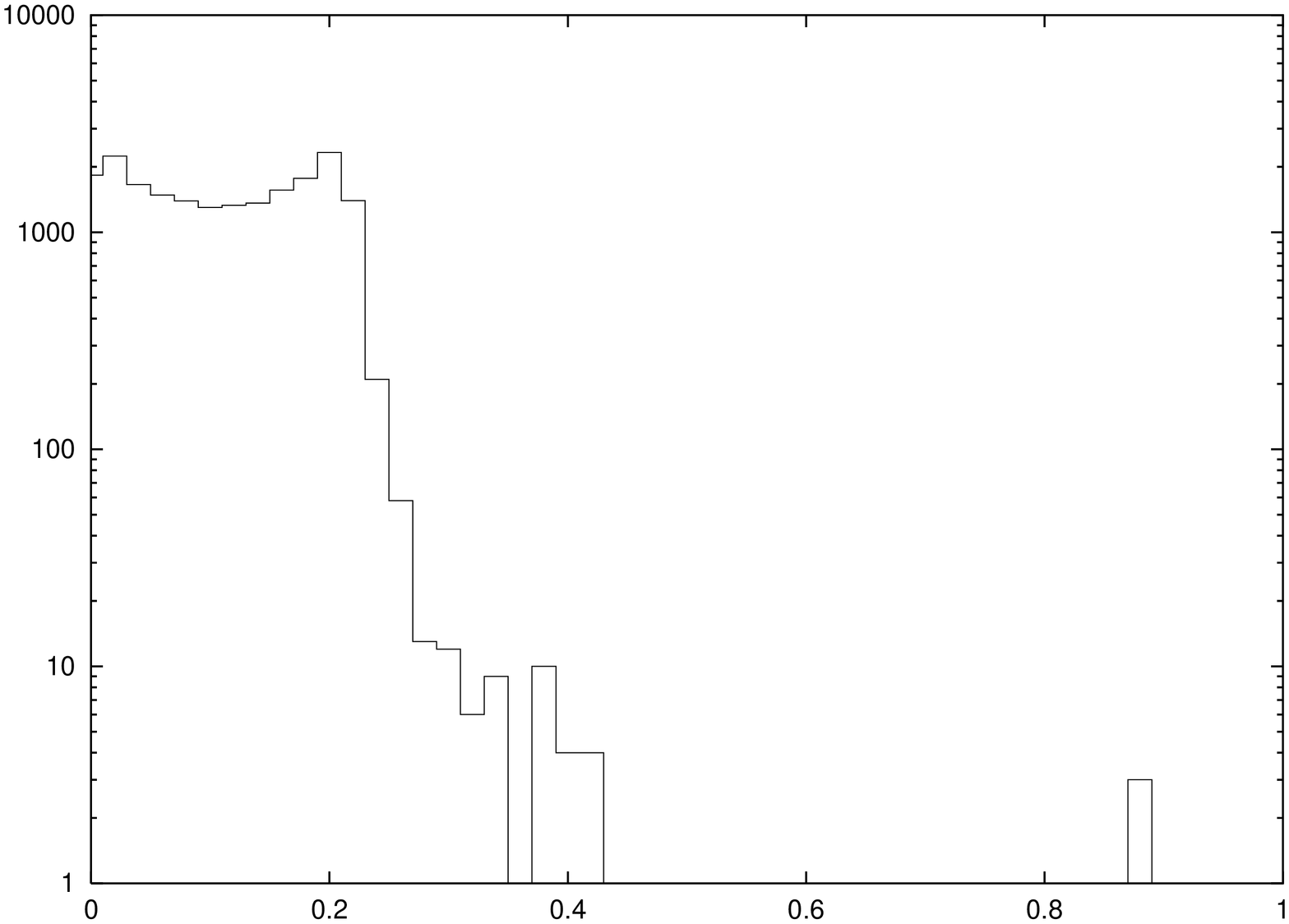, height = 16.0cm, width = 7.0cm, angle = 270}}
 \mbox{\epsfig{file = 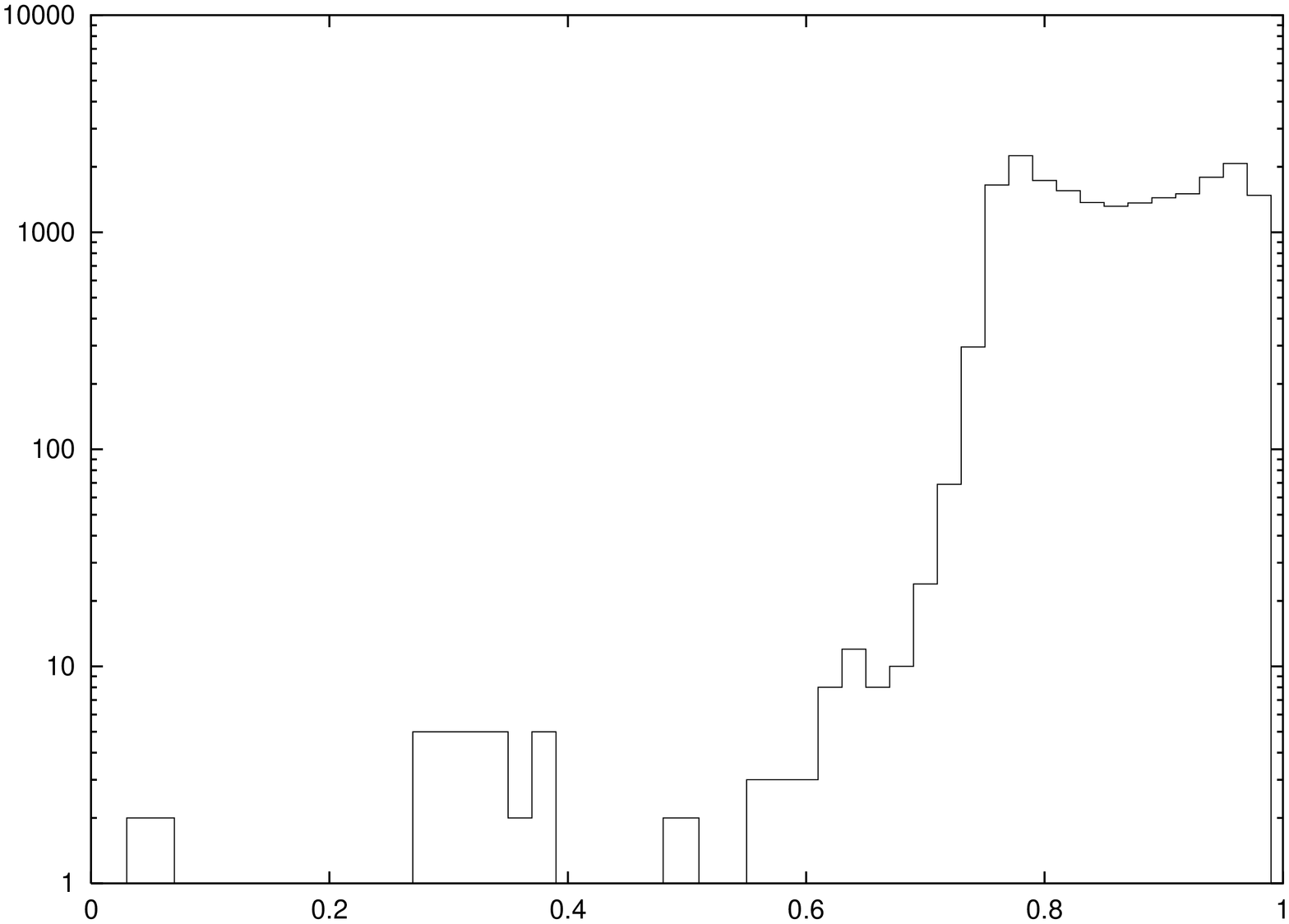, height = 16.0cm, width = 7.0cm, angle = 270}}
 \mbox{\epsfig{file = 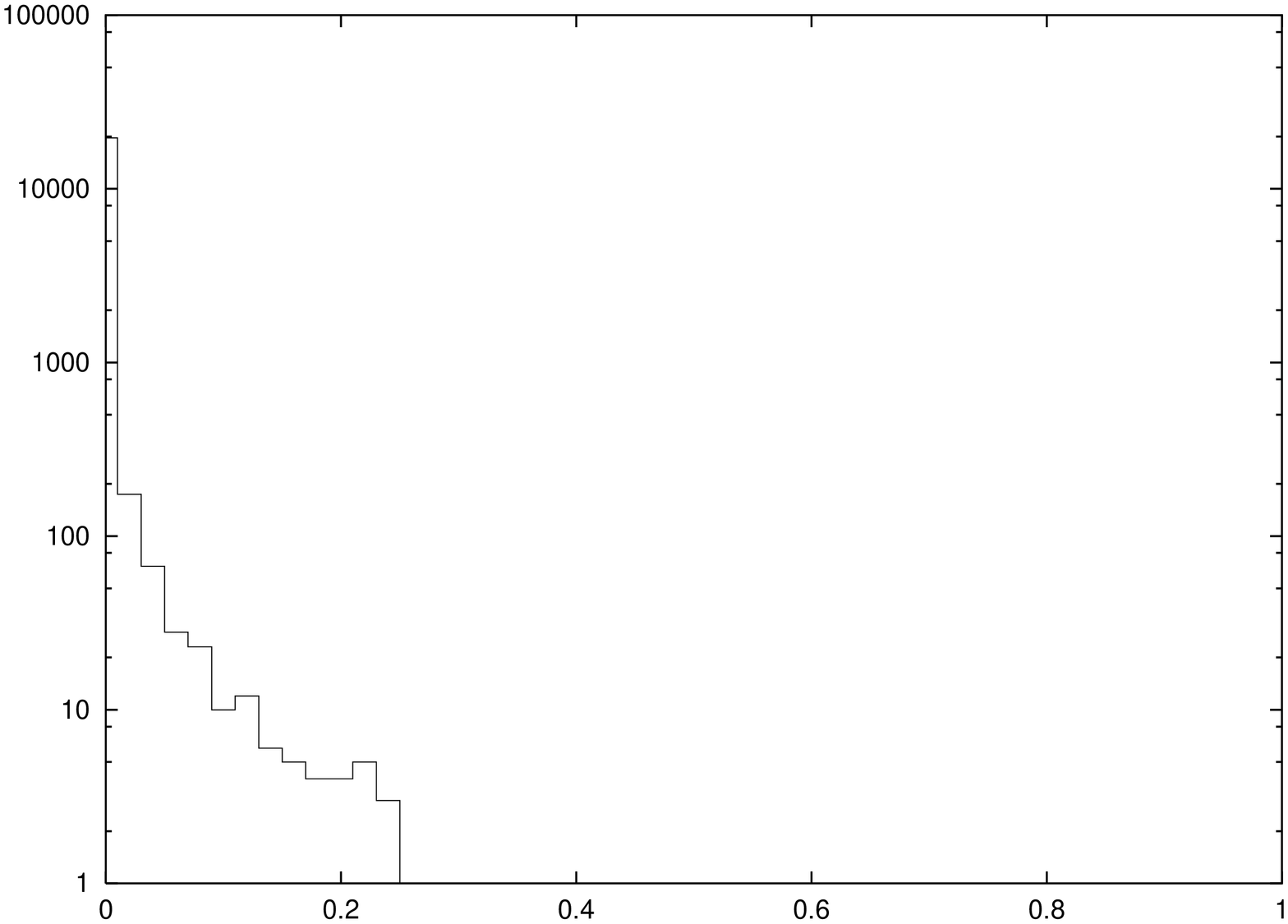, height = 16.0cm, width = 7.0cm, angle = 270}}
  \protect
 \caption{\it Diagonal matrix elements: $h_{--}, h_{00}, h_{++}$ respectively for $\rho=0.09$, 
   $\eta=0.323$. }
  \end{center}
\end{figure}     
%%%
%%%%%%%%%%
%%%%%%%%%%%%%%%%%%%%%%%%%%   VALEURS NON-DIAGONALES %%%%%%%%%%%%%%%%%%%%%%%%%%%%%%%%%%%%%%%%%%%%%%%%
%%%
\begin{figure}[htbp]
  \begin{center}
 \mbox{\epsfig{file = 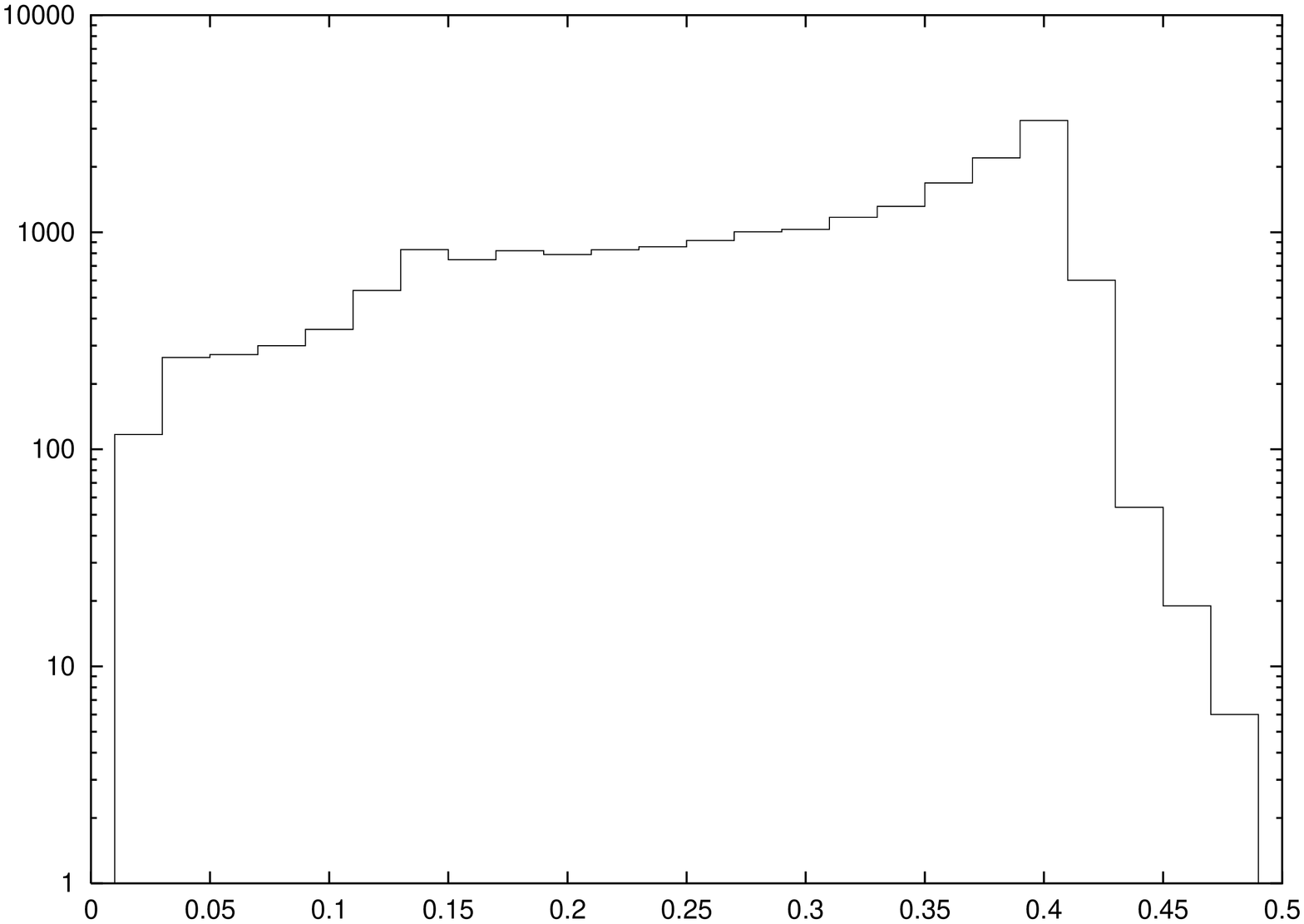, height = 16.0cm, width = 7.0cm, angle = 270}}
 \mbox{\epsfig{file = 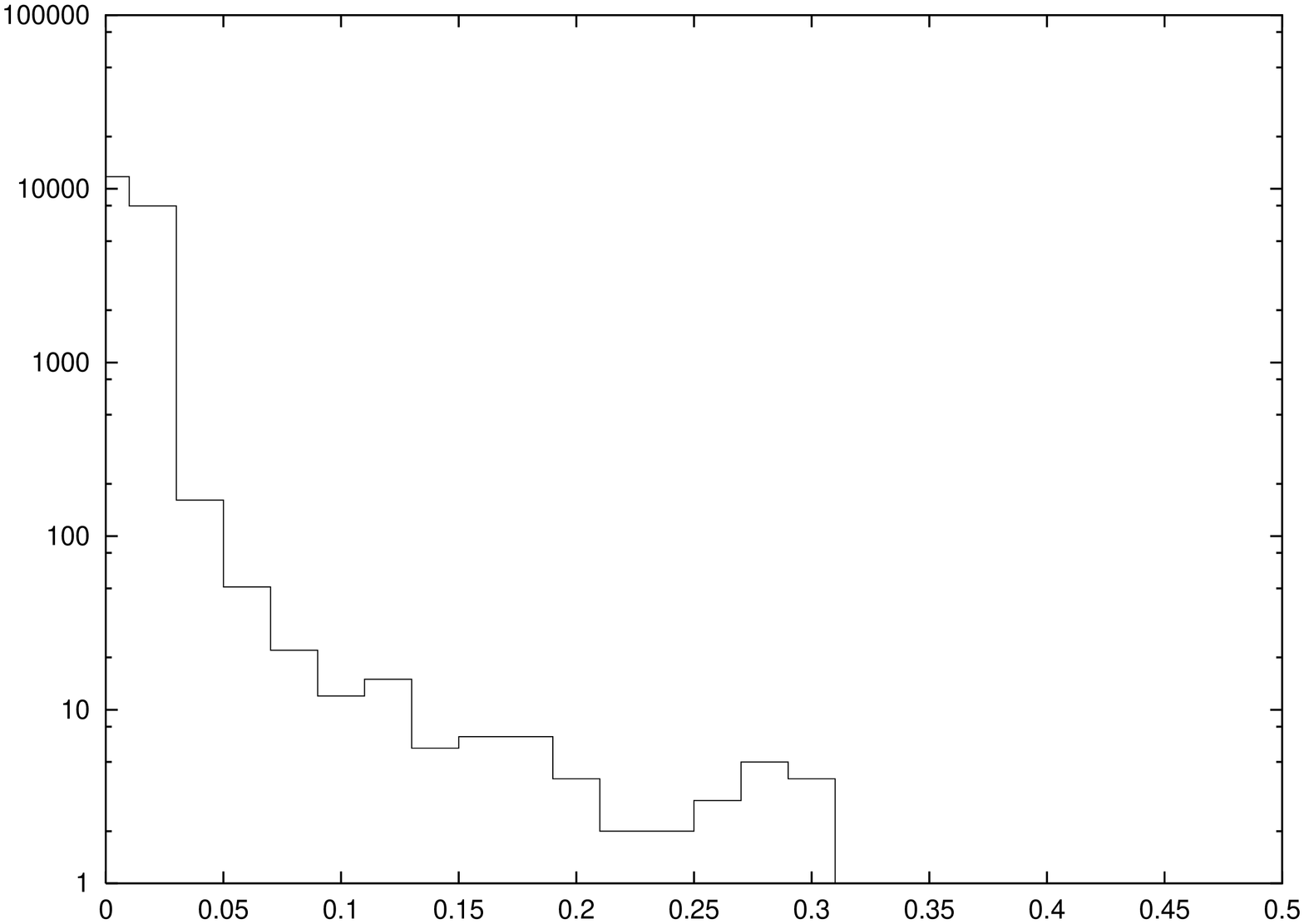, height = 16.0cm, width = 7.0cm, angle = 270}}
 \mbox{\epsfig{file = 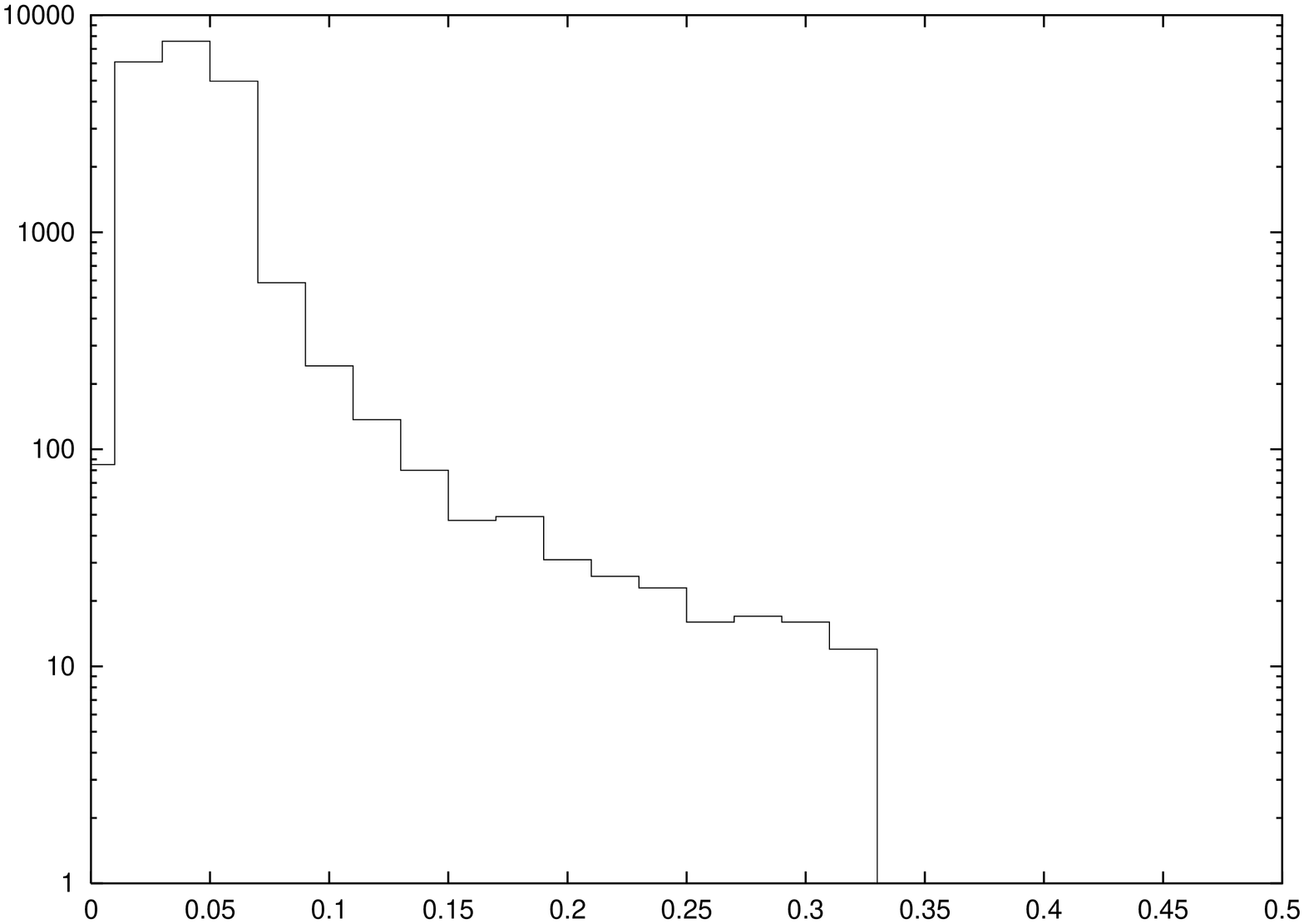, height = 16.0cm, width = 7.0cm, angle = 270}}
 \protect
\caption{\it Modulus of non diagonal matrix elements: $h_{0-}, h_{+-}, h_{+0}$ respectively
 for $\rho = 0.09, \eta = 0.323$. }
  \end{center}
\end{figure}     
%%%%%%%%%%%%%%%%%%%%%%%%%%%%%%%%%%%%%%%%%%%%%%%%%%%%%%%%%%%%%%%%%%%%%%%%%%%%%%%%%%%%%%%%%%%%%%%%%%%%%%%%%%%

%%%%%%%%%%%%%%%%%%%%%%%%%%%%%%%%%% hii pour differentes valeurs de rho et eta (Wolfenstein param)%%%%%%%
\begin{figure}[htbp]
  \begin{center}
 \mbox{\epsfig{file = 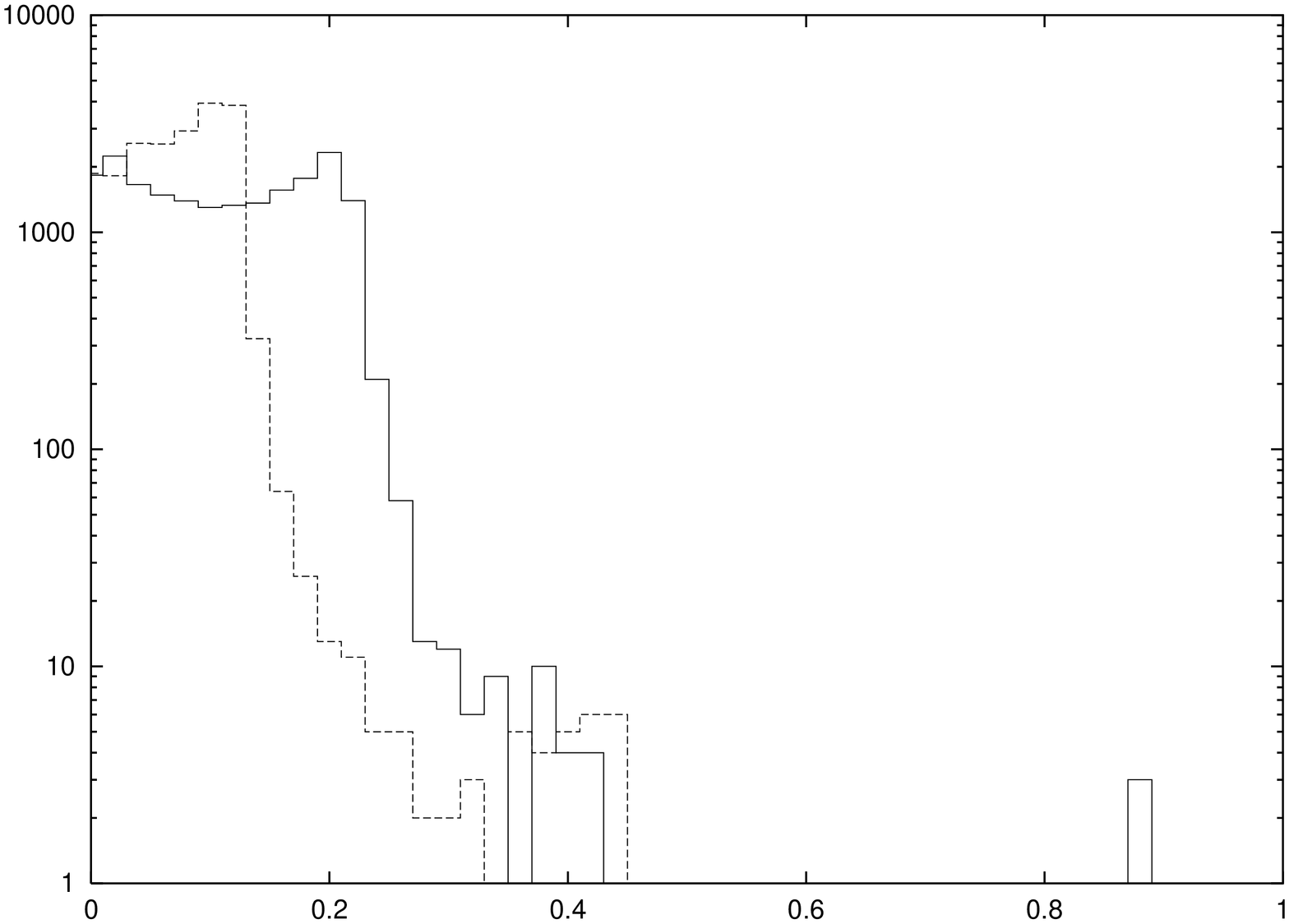, height = 16.0cm, width = 7.0cm, angle = 270}}
 \mbox{\epsfig{file = 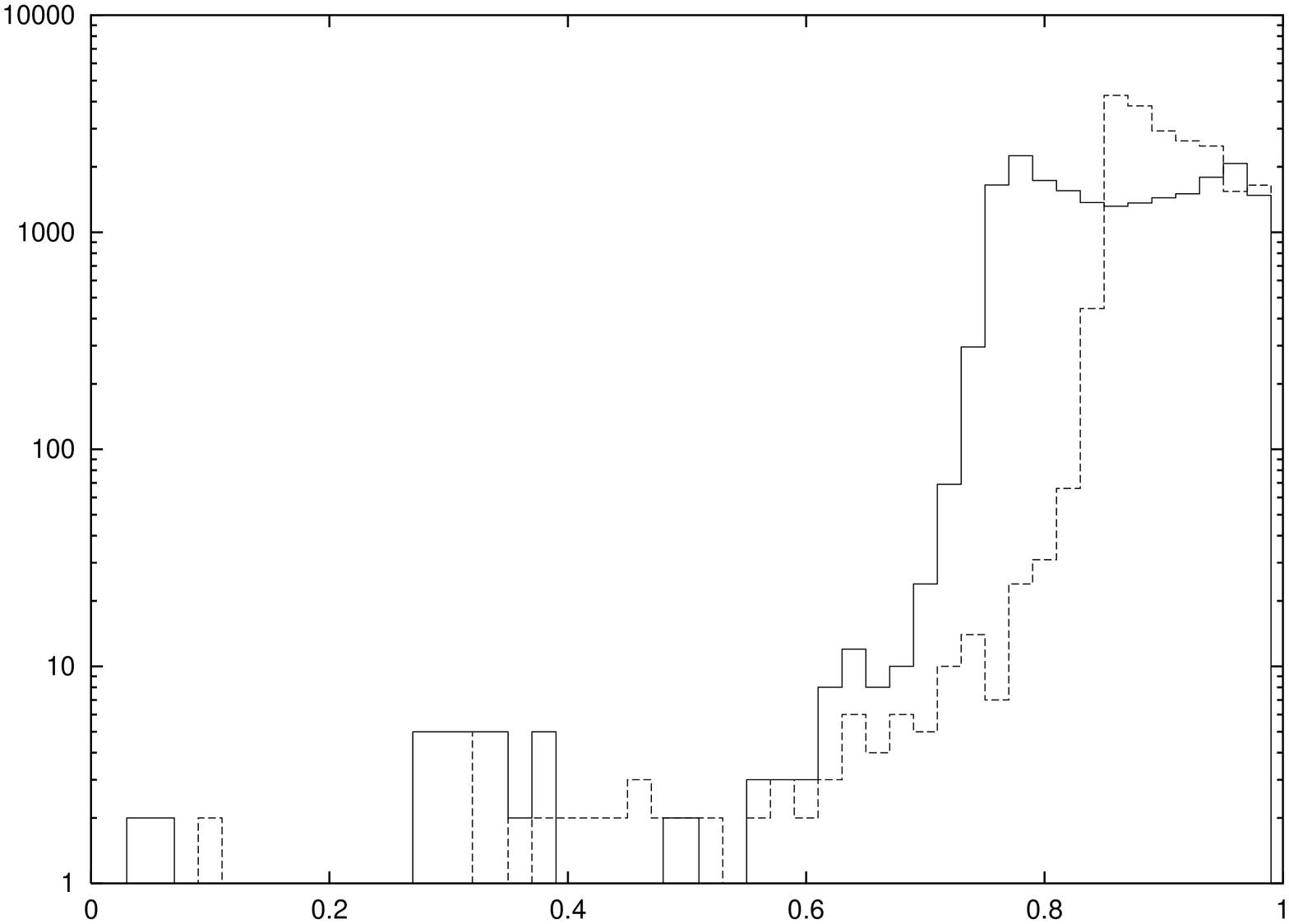, height = 16.0cm, width = 7.0cm, angle = 270}}
 \mbox{\epsfig{file = 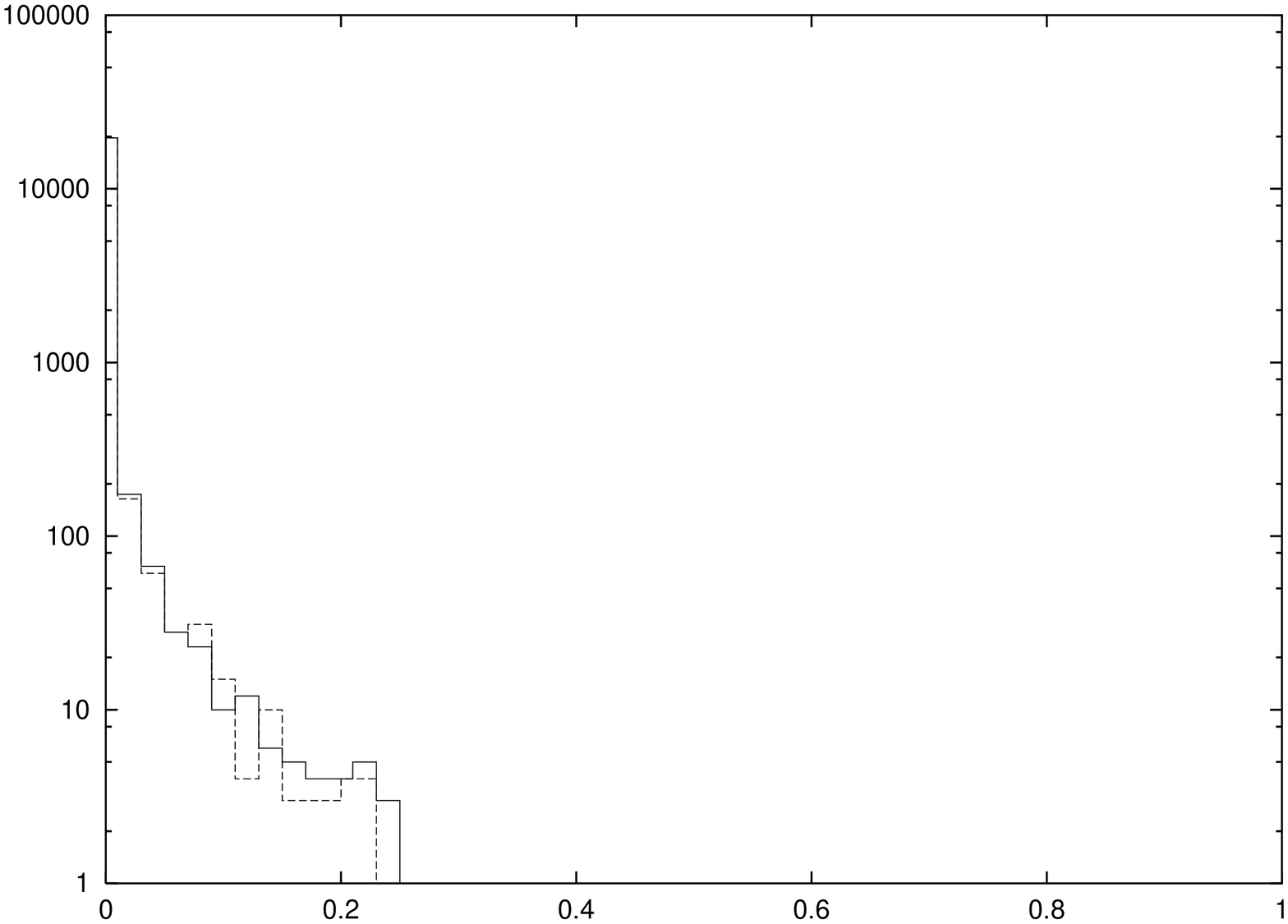, height = 16.0cm, width = 7.0cm, angle = 270}}
 \protect
 \caption{\it Variations of $h_{--}, h_{00}, h_{++}$ according to Wolfenstein parameters: 
  $\rho = 0.09, \eta = 0.323$ (full line) and $\eta = 0.442$ (dashed line).}
  \end{center}
\end{figure}     
%%%%%%%%%%%%%%%%%%%%%%%%%%%%%%%%%%%%%%%%%%%%%%%%%%%%%%%%%%%%%%%%%%%%%%%%%%%%%%%%%%%%%%%%%%%%%%%%%%%%%%%%%%%
\begin{figure}[htbp]
  \begin{center}
 \mbox{\epsfig{file = 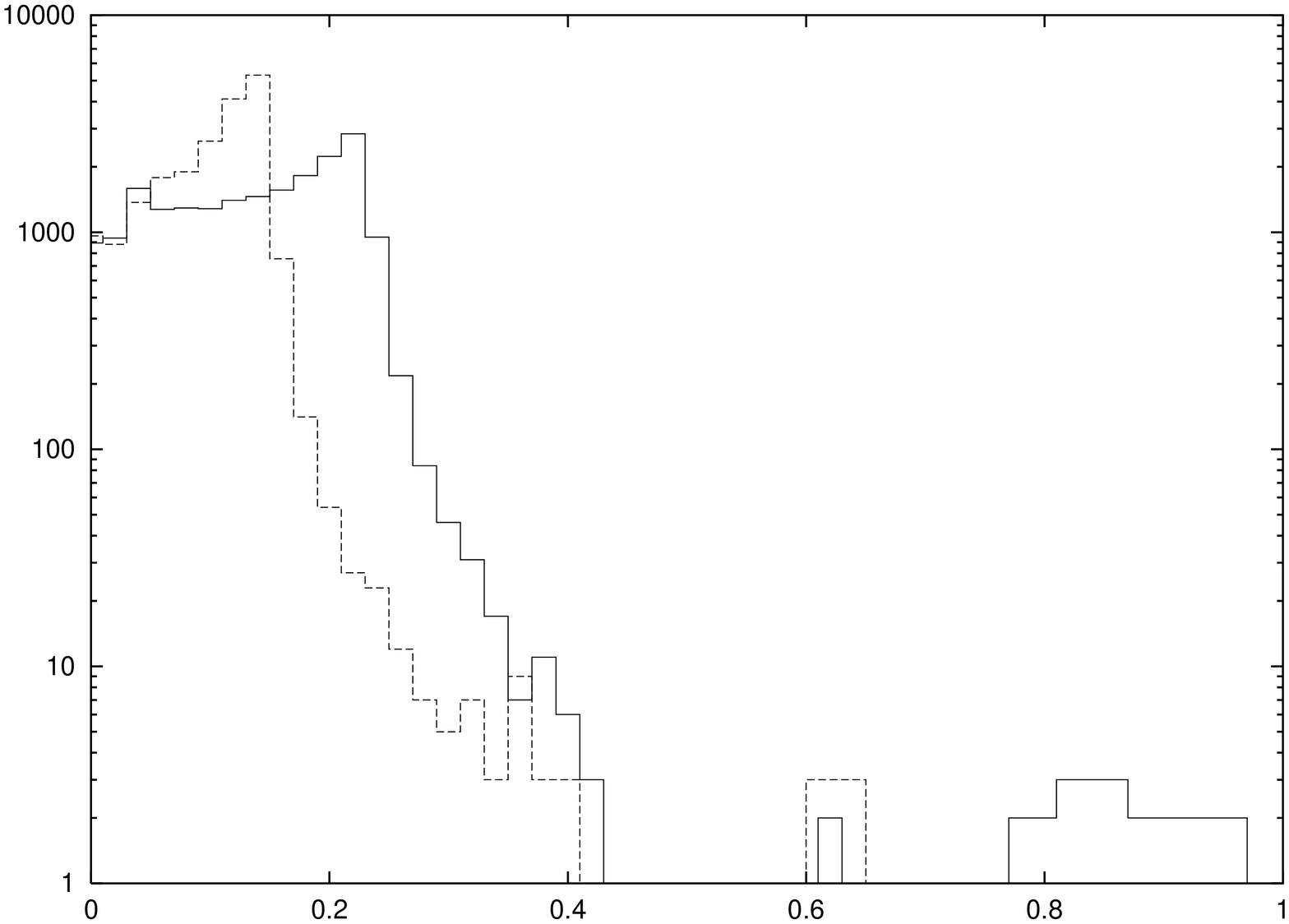, height = 16.0cm, width = 7.0cm, angle = 270}}
 \mbox{\epsfig{file = 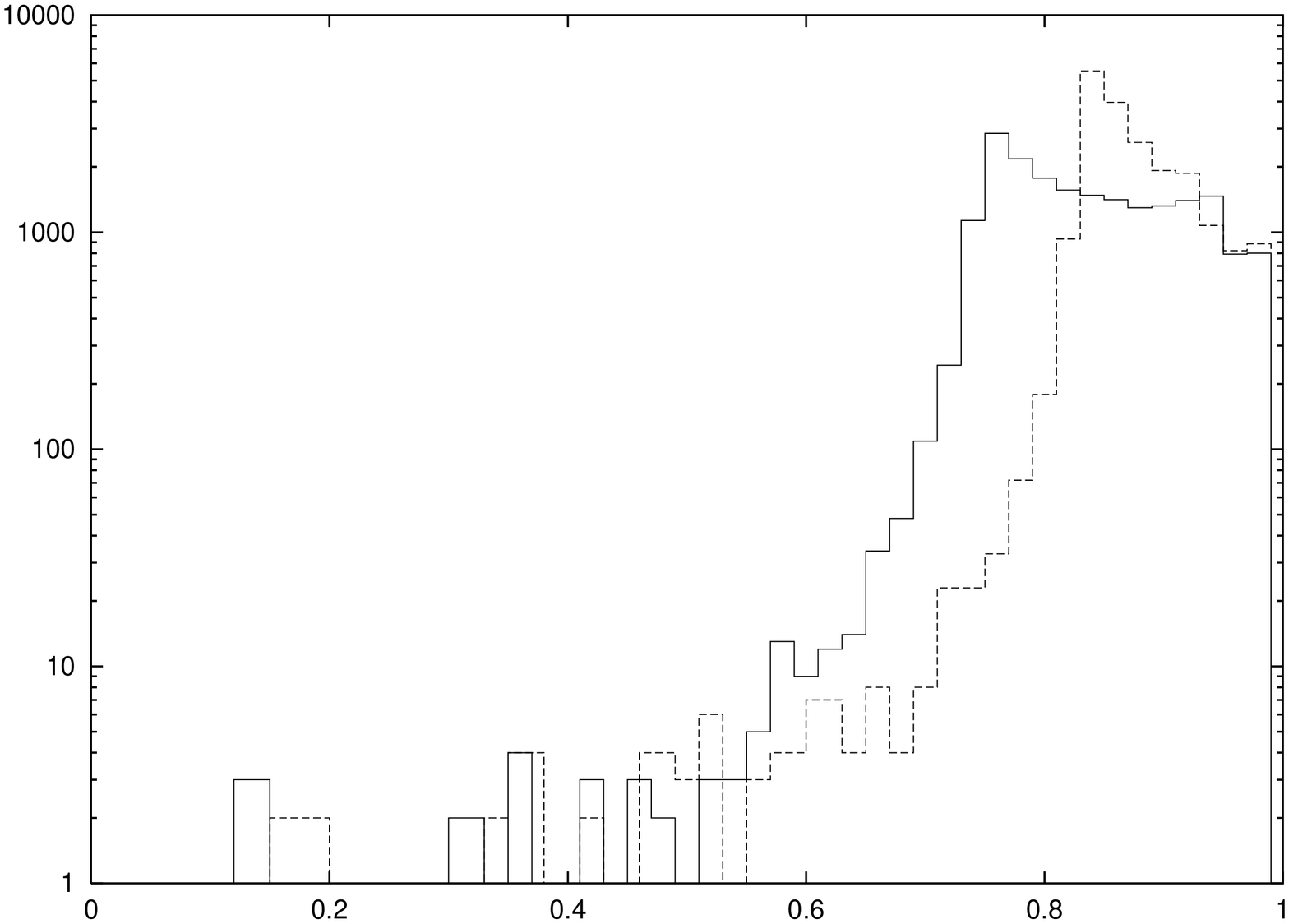, height = 16.0cm, width = 7.0cm, angle = 270}}
 \mbox{\epsfig{file = 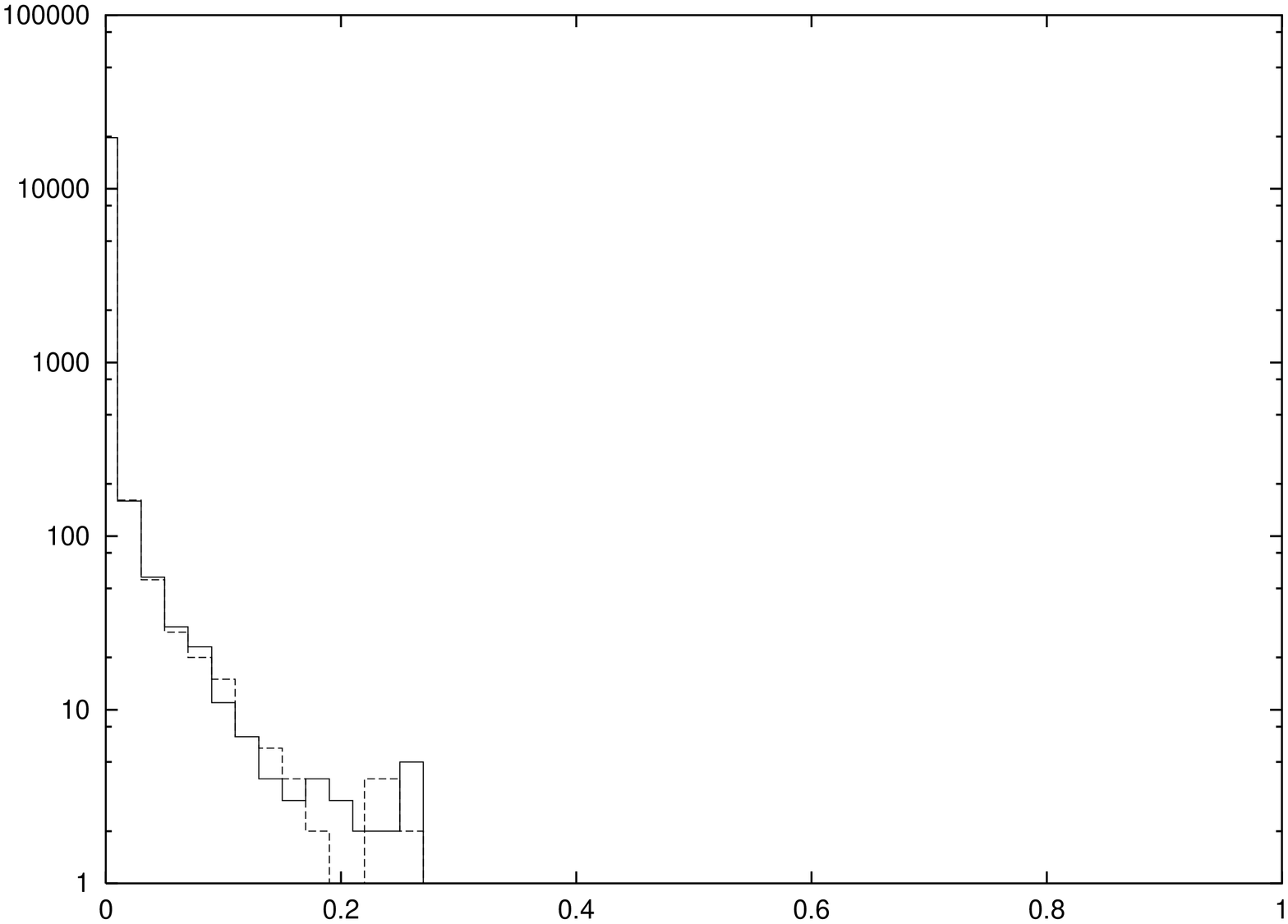, height = 16.0cm, width = 7.0cm, angle = 270}}
 \protect
 \caption{\it Variations of $h_{--}, h_{00}, h_{++}$ according to Wolfenstein parameters:
  $\rho = 0.254, \eta = 0.323$ (full line) and $\eta = 0.442$ (dashed line).}
  \end{center}
\end{figure}     
%%%%%%%%%%%%%%%%%%%%%%%%%%%%%%%%%%%%%%%%%%%%%%%%%%%%%%%%%%%%%%%%%%%%%%%%%%%%%%%%%%%%%%%%%%%%%%%%%%%%%%%%%%%

%%%%%%%%%%%%%%%%
%%%%%%%%%%%%%%%%% PARTIES REELLE ET IMAGINAIRE DES ELEMENTS NON-DIAGONAUX %%%%%%%%%%%%%%%%%%%
%%%
\begin{figure}[htbp]
  \begin{center}
 \mbox{\epsfig{file = 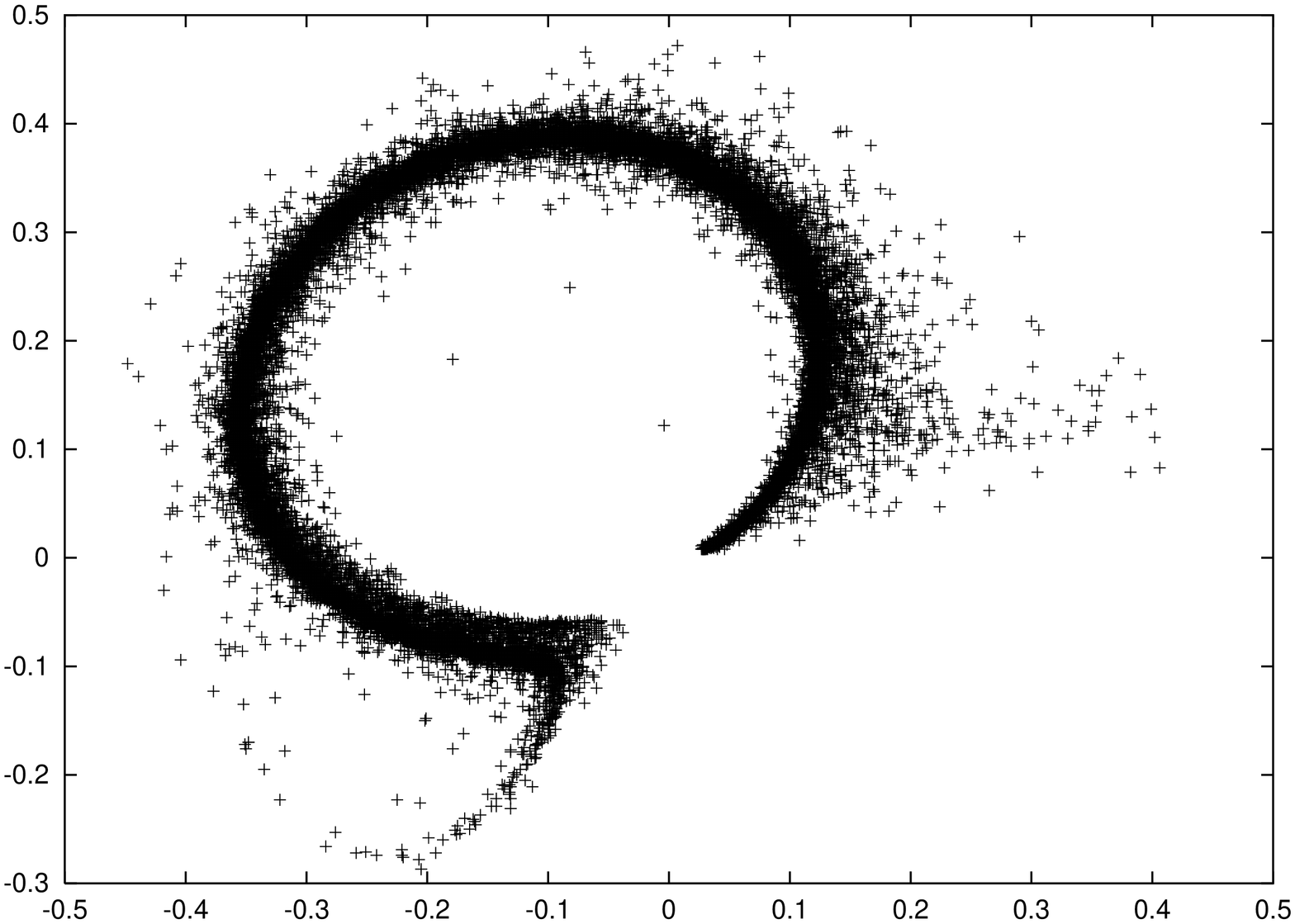, height = 16.0cm, width = 7.0cm, angle = 270}}
 \mbox{\epsfig{file = 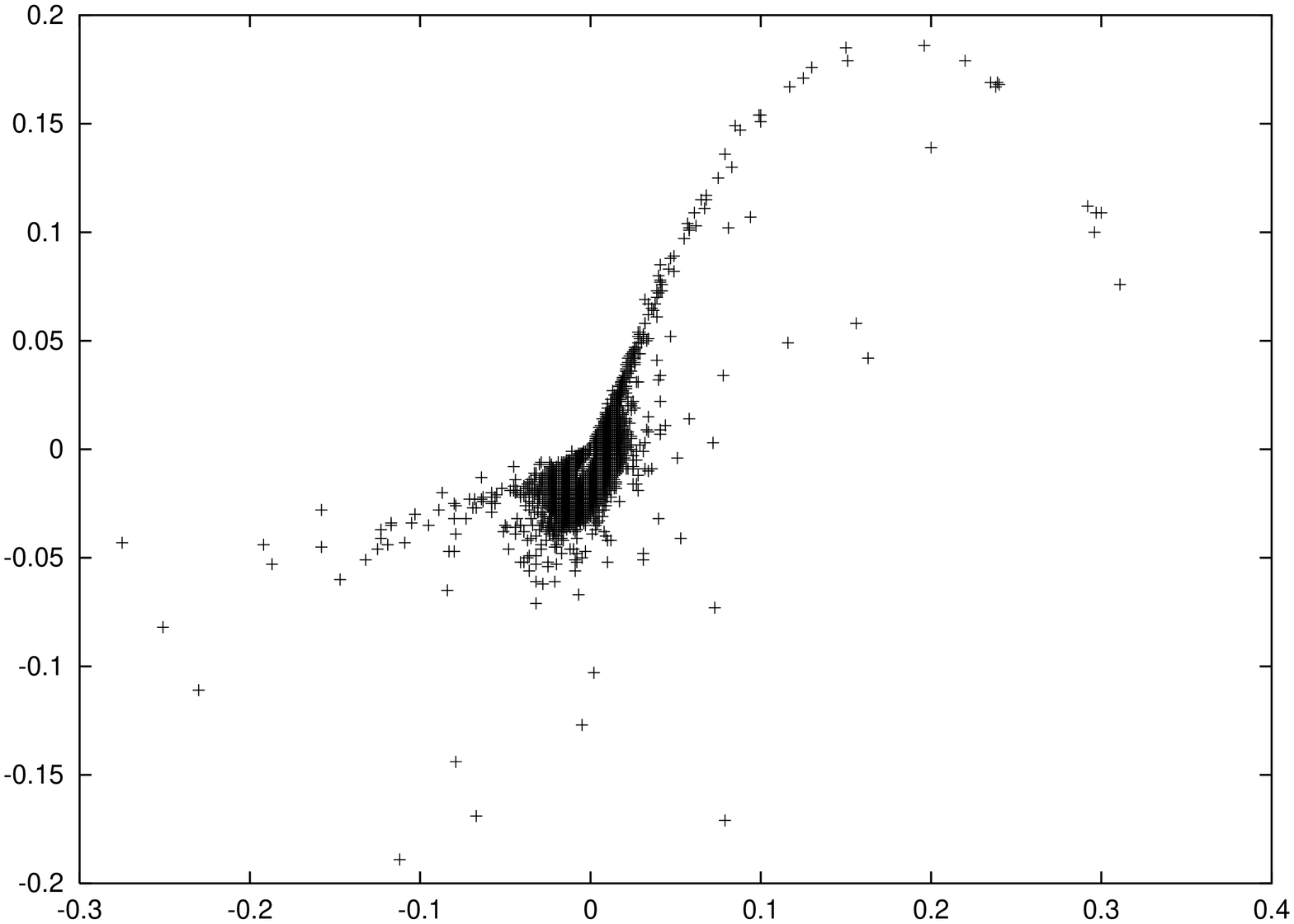, height = 16.0cm, width = 7.0cm, angle = 270}}
 \mbox{\epsfig{file = 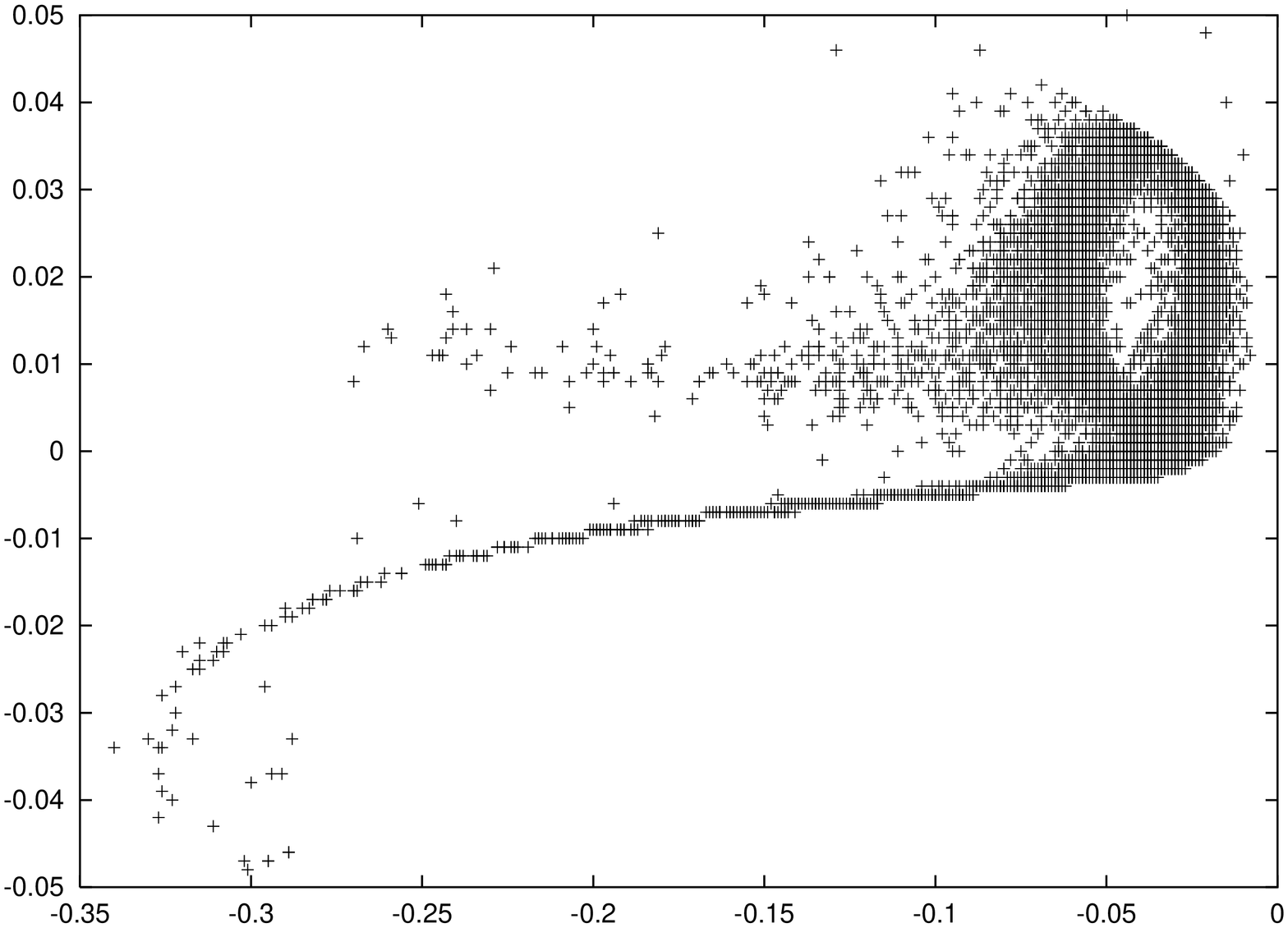, height = 16.0cm, width = 7.0cm, angle = 270}}
 \protect
\caption{\it Imaginary part vs Real part of matrix elements $h_{0-}, h_{+-}, h_{+0}$ 
         respectively for $\rho = 0.09, \eta = 0.323$.}
\end{center}
\end{figure}     
%%%%%%%%%%%%%%%%%%%%%%%%%%%%%%%%%%%%%%%%%%%%%%%%%%%%%%%%%%%%%%%%%%%%%%%%%%%%%%%%%%%%%%%%%%%%%%%%%%%%%%%%%%%
%%%
\begin{figure}[htbp]
  \begin{center}
 \mbox{\epsfig{file = 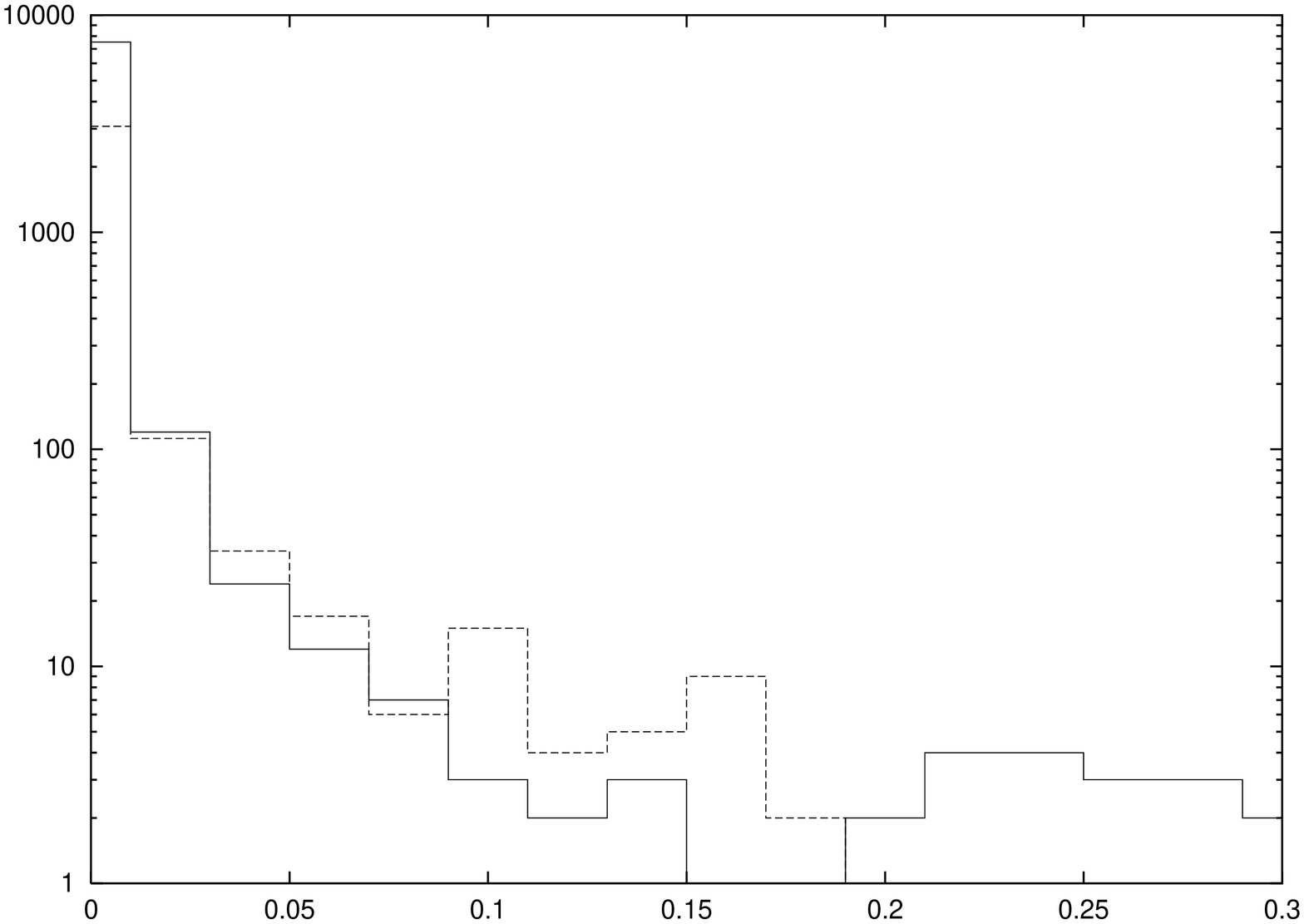, height = 16.0cm, width = 5.5cm, angle = 270}}
 \mbox{\epsfig{file = 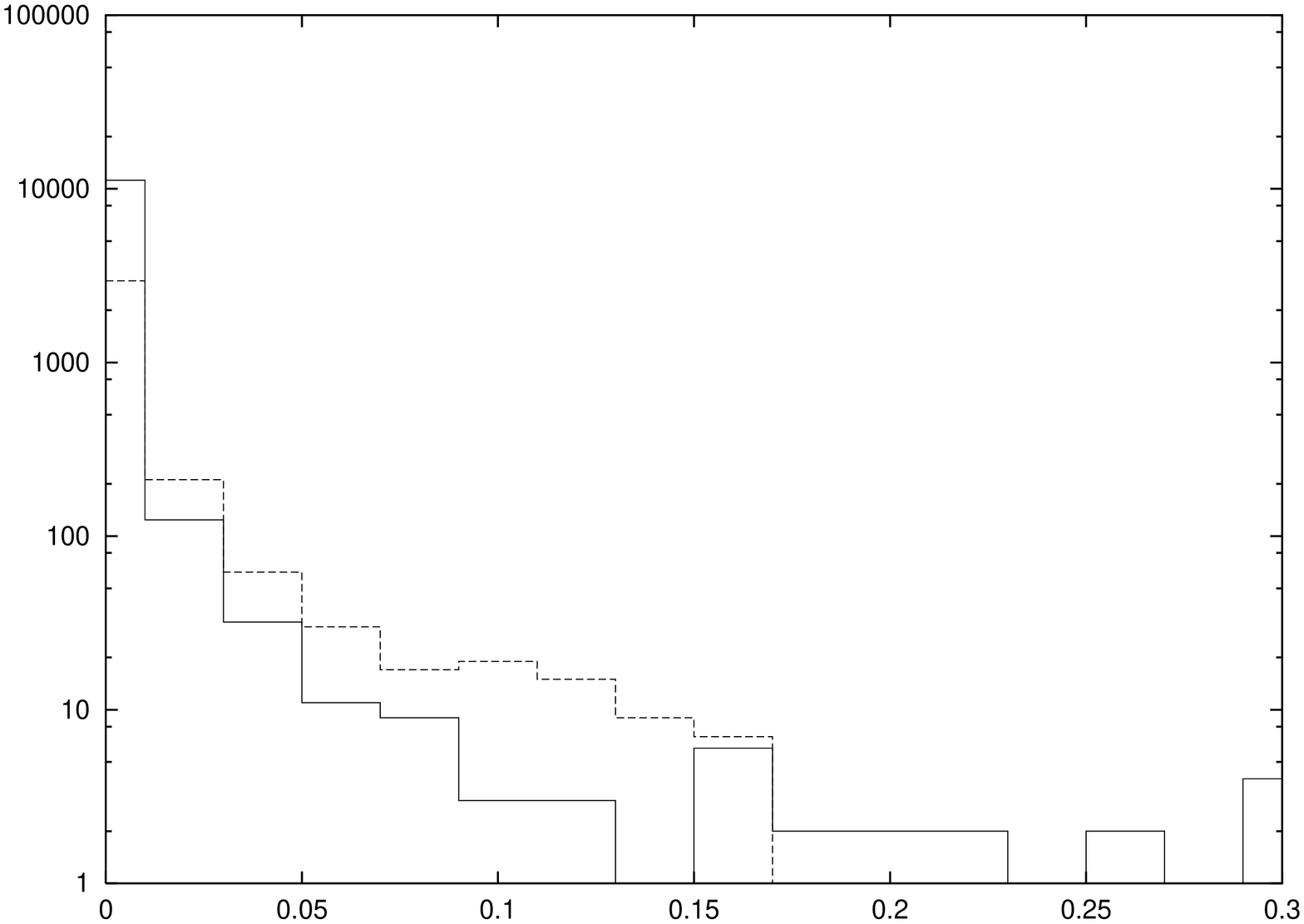, height = 16.0cm, width = 5.5cm, angle = 270}}
 \mbox{\epsfig{file = 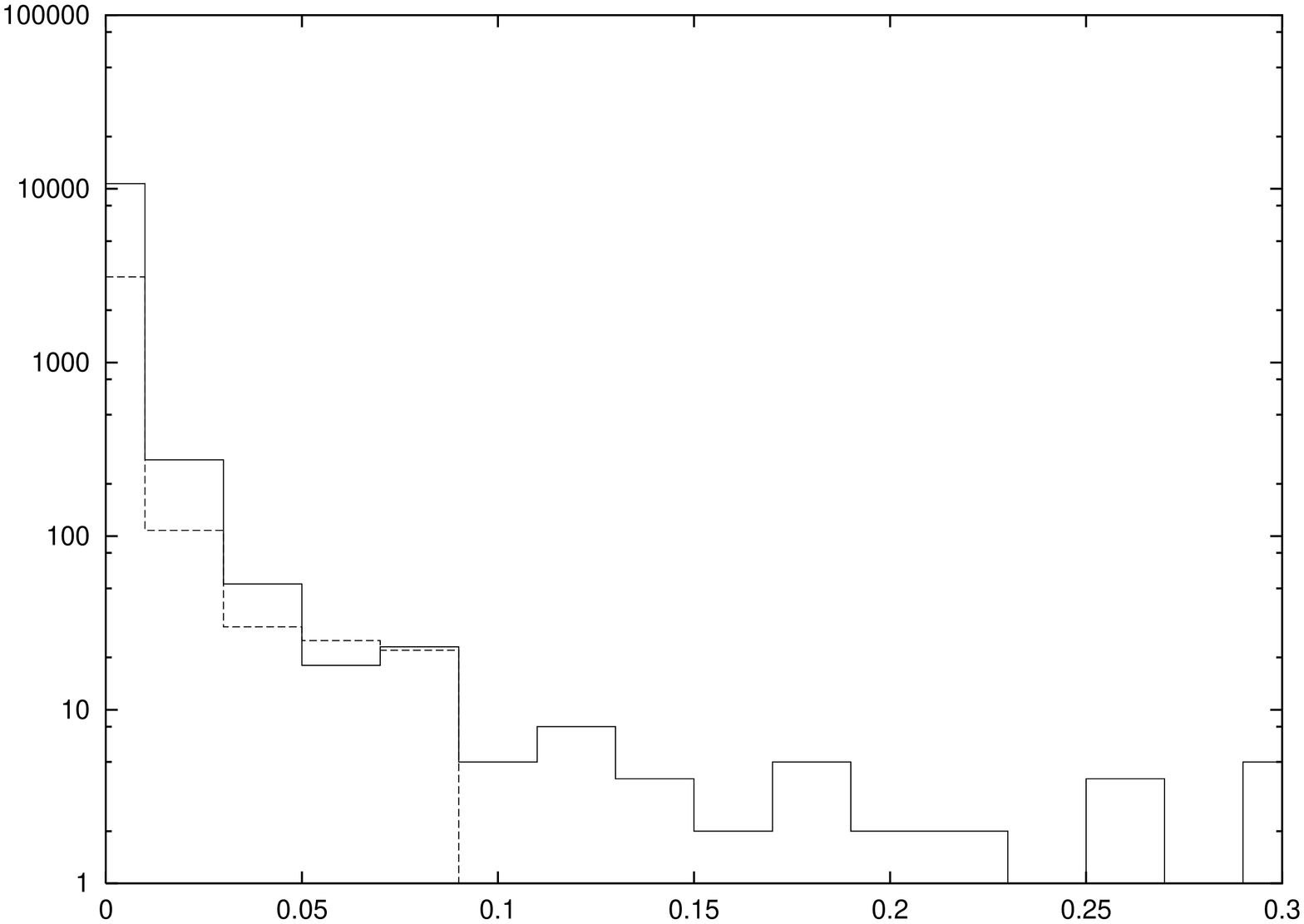, height = 16.0cm, width = 5.5cm, angle = 270}}
 \mbox{\epsfig{file = 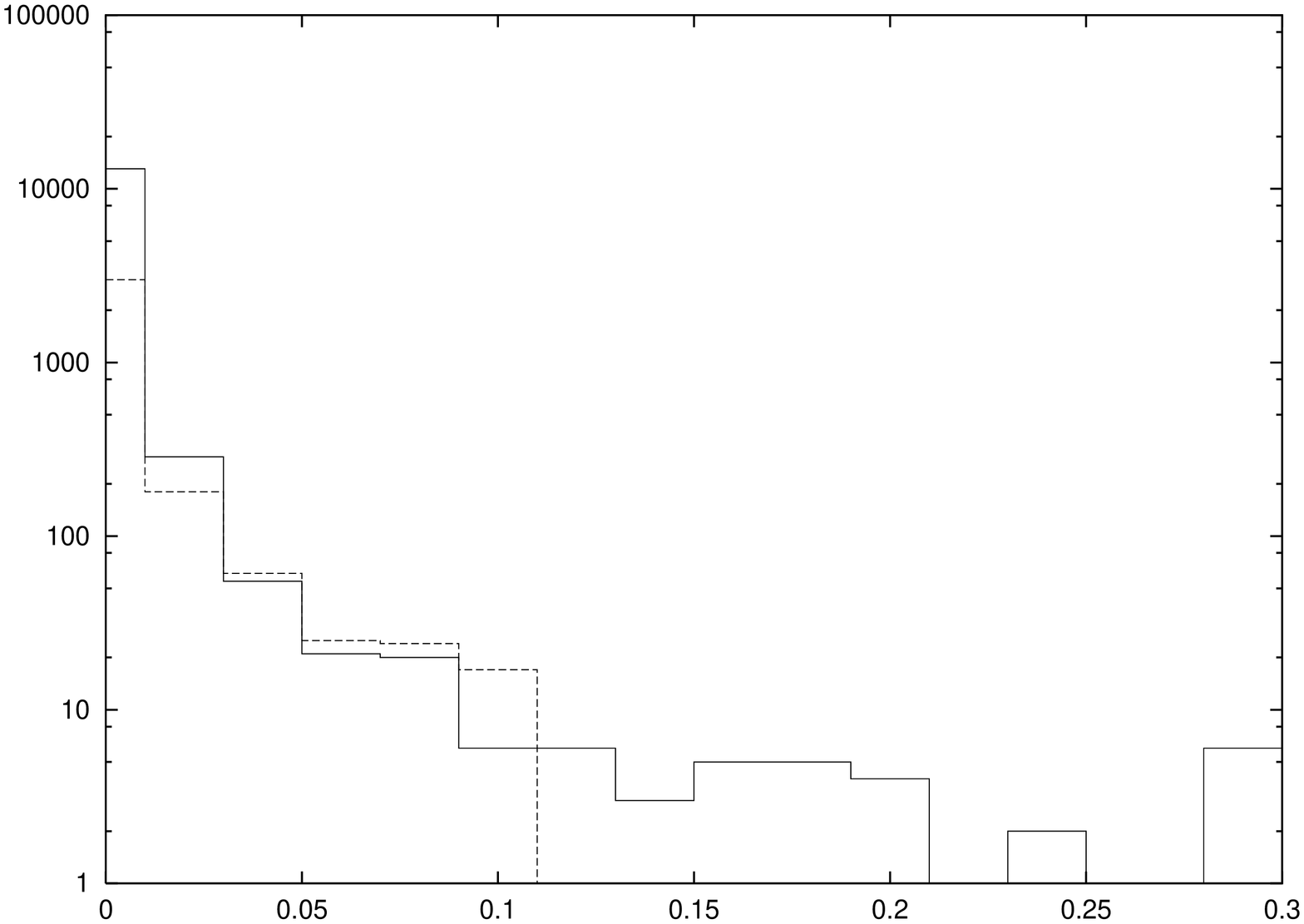, height = 16.0cm, width = 5.5cm, angle = 270}}
 \protect
\caption{\it Real part (full line) and Imaginary part (dashed line) of $h_{+-}$ matrix element 
  for $(\rho, \eta)$ = (0.09, 0.323); (0.09, 0.442); (0.254, 0.323) and (0.254, 0.442) respectively.}
%%%  $\rho = 0.09 \ , \ \eta = 0.323$ ; $\rho = 0.09 \ , \ \eta = 0.442$ ; $\rho = 0.254 \ , \ \eta = 0.323$ 
%%%  and $\rho = 0.254 \ , \eta = 0.442$ respectively}
\end{center}
\end{figure}     
%%%%%%%%%%%%%%%%%%%%%%%%%%%%%%%%%%%%%%%%%%%%%%%%%%%%%%%%%%%%%%%%%%%%%%%%%%%%%%%%%%%%%%%%%%%%%%%%%%%%%%%%%%%

\newpage

\section{Decays of vector mesons $V_1 \  V_2$ into two pseudoscalar mesons}

  The matrix elements derived above allow us to compute the {\it degrees of polarization} of each resonance
like $K^*$ or ${\rho}^0$. The angular distributions of the pseudoscalar mesons in each $V_i$ rest frame depend
on: \\
$(i)$ the spin $1$ of the vector meson $V_i$.  \\
$(ii)$ the weight of each helicity state.   \\
$(iii)$ the correlations among the helicity states of the two vector mesons. 

\vskip 0.5cm

Complete analytical expression of the final angular distributions is  the following one:
  
%%%%%%%%%%%%%%%%%%%%%%%%%%%%%%%%%%%% V1----> l+ l- (J/Psi) %%%%%%%%%%%%%%%%%%%%%%%%%%%%%%%%  
  
 %%\begin{eqnarray*}
%%%{{d^3\Gamma}\over{d\cos\theta_1 d\cos\theta_2 d\phi}} & \propto &
%%%\left({(|H_+|^2 +|H_-|^2) {\sin}^2{\theta_1}(1+{\cos}^2\theta_2) + 
%%%{|H_0|^2}{\cos}^2{\theta_1} {\sin}^2 {\theta_2} } \right.\\ 
%%% & &  {
%%%- 2{\sin}^2{\theta_2} {\sin}^2{\theta_1} {(Re(H_+ H^*_-){\cos{2\phi}} - 
%%% Im(H_+ H^*_-){\sin{2\phi}})}  } \\
%% & & \left. {
%% -  {\sin{2\theta_1}} {\sin{2\theta_2}}
%%{({Re(H_+ H^*_0 + H_- H^*_0)}{\cos\phi} - {Im(H_+ H^*_0 - H_- H^*_0)}
%%{\sin\phi})}} \right)
%%\end{eqnarray*}
%%%%%%%%%%%%%%%%%%%%%%%%%%%%%%%%%%%%%%%%%%%%%%%%%%%%%%%%%%%%%%%%%%%%%%%%%%%%%%%%%%%%%%%%%%%%%%%%%%%%%%%%%%

\begin{eqnarray}
{{d^3\Gamma}\over{d\cos\theta_1 d\cos\theta_2 d\phi}} & \propto &
{(h_{++} + h_{--})}{{\sin}^2{\theta_1}{\sin}^2{\theta_2}}/4 + {h_{00}{\cos}^2{\theta_1}{\cos}^2{\theta_2}} \nonumber
\\
& &\!\!\! \!\!\!\!\!\!\!\! \!\!\!\!\! \!\!\!\!\!\!\!\! \!\! \!\!\!\!\!\!\!\! \!\!\!\!\!   + \left(\Re e{(h_{+0})}{\cos{\phi}} - \Im m{(h_{+0})}{\sin{\phi}} + \Re e{(h_{0-})}{\cos{\phi}} -
\Im m{(h_{0-})}{\sin{\phi}}\right){{\sin{2\theta_1}}{\sin{2\theta_2}}}/4  \nonumber \\
& & \!\!\!\!\!\! \!\! \!\!\!\!\!\!\!\! \!\!\!\!\!\!\!\! \!\!\!\!\!\!\!\!\!\! \!\!\!\!\!   + 
\left(\Re e{(h_{+-})}{\cos{2\phi}} - \Im m{(h_{+-})}{\sin{2\phi}}\right){{\sin}^2{\theta_1}{\sin}^2{\theta_2}}/2.
\end{eqnarray}

Angles ${\theta}_1$ , ${\theta}_2$ and $\phi$ have been defined in Section $1$. 
%%are the polar angles of particles $a_1$ and $a_2$ in the
%%$V_1$ and $V_2$ rest frames ; while angle $\phi$ is given by ${\phi}_2 - {\phi}_1 \ $,  where ${\phi}_i$ 
%%is the corresponding azimuthal angle of $a_i$ in $V_i$ rest frame.

\vskip 0.5cm
Explicit angular distributions for polar and azimuthal angles can be derived from the relation above.
 It is interesting to notice that,
 due to the pseudoscalar nature of the final particles, angles $\theta_1$ and
$\theta_2$ have the {\it same} distributions:

 \vskip 0.4cm
\begin{itemize}
\item  ${d{\sigma}}/{d{\cos{\theta_{1, 2}}}} \propto (3h_{00}-1)\cos^{2}{\theta_{1, 2}} + (1-h_{00}).$
\item ${d{\sigma}}/{d{\phi}} \propto (1+2(\Re e {(h_{+-})}\cos{2\phi} - \Im m{(h_{+-})}\sin{2\phi})).$
\end{itemize}

 %%%%%%%%%%%%%%%%%% ANGLES POLAIRE et AZIMUTAL %%%%%%%%%%%%%%%%%%%%%%%%%%%%%%%%%%%
 
\begin{figure}[htbp]
  \begin{center}
 \mbox{\epsfig{file = 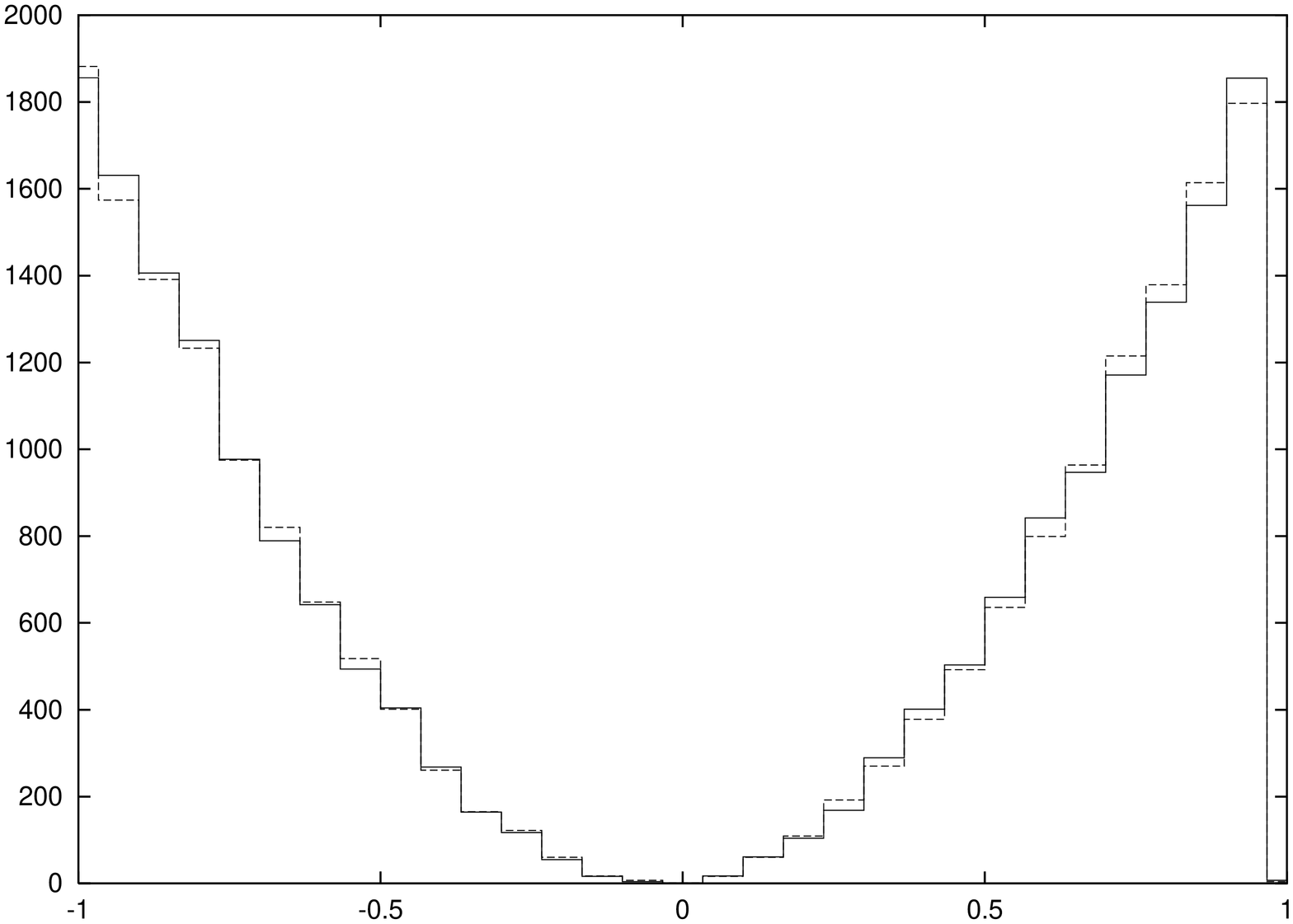, height = 15.0cm, width = 10.0cm, angle = 270}}
 \mbox{\epsfig{file = 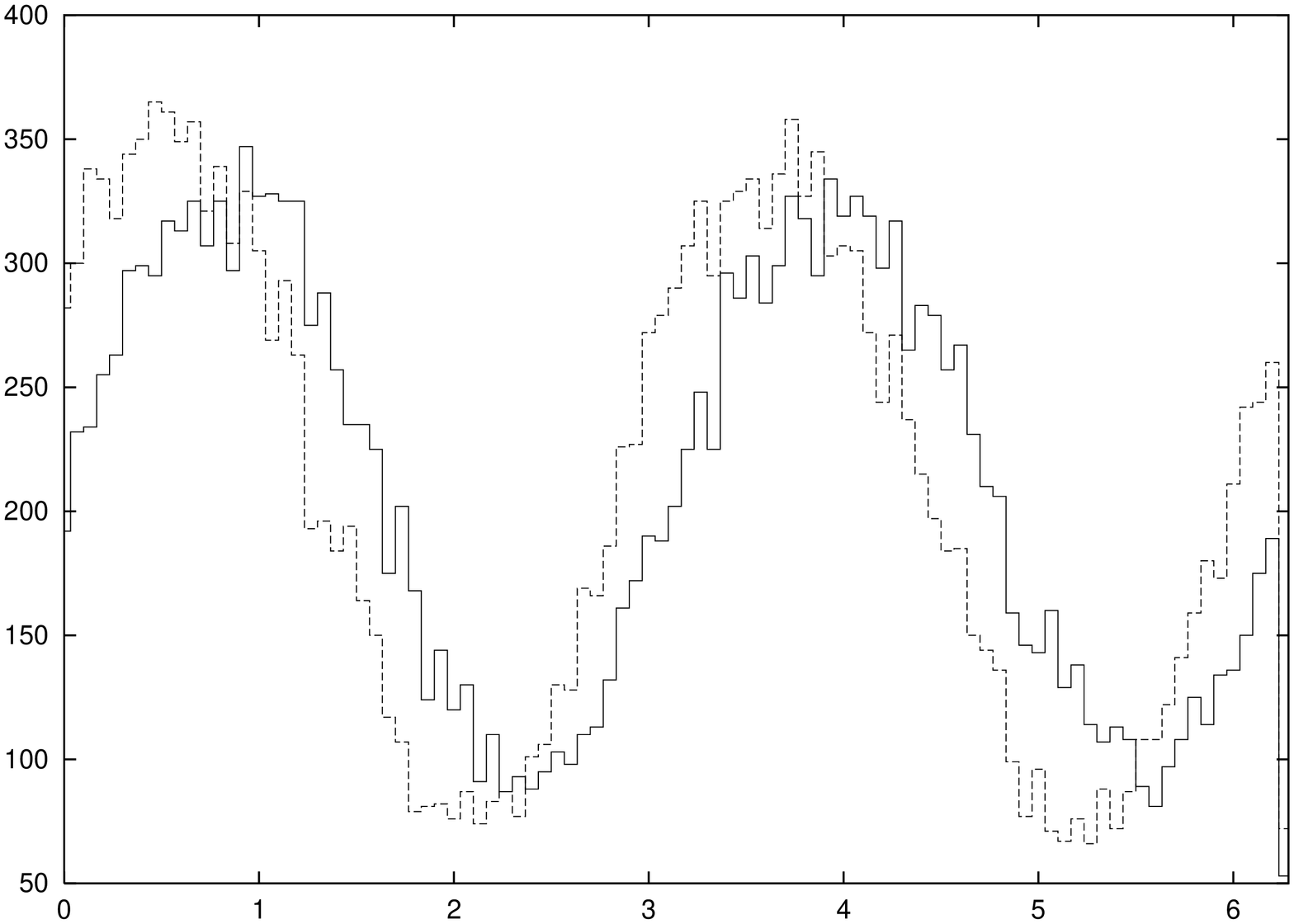, height = 15.0cm, width = 10.0cm, angle = 270}}
 \protect
\caption{\it Cos${\theta}$ distribution (upper figure) and azimuthal angle $\phi$ distribution 
(lower figure) for $\rho = 0.09, {\eta} = 0.323$ (full line) and $\rho = 0.254, {\eta} = 0.442$ 
 (dashed line) respectively. }
  \end{center}
\end{figure}

%%%%%%%%%%%%%%%%% COMMENTAIRES SUR LES DISTRIBUTIONS ANGULAIRES %%%%%%%%%%%%%%%%%%%%%%
\vskip 0.4cm

In Fig.10 are displayed respectively the $\cos{\theta}$ distribution and the azimuthal angle $\phi$ one.
 Some comments on these curves are necessary:
 
\begin{itemize} 
 \item  The $\cos{\theta}$ distribution is practically the same
 whatever the values of $\rho$ and $\eta$ are; no sensitivity to particular values of $\rho$ and $\eta$
is seen. 
\item As far as angle $\phi$ is concerned, its distribution depends on the matrix element $h_{+-}$. Despite 
the fact that $\Re e {(h_{+-})}$ and $\Im m{(h_{+-})}$ do not exhibit sensitive differences (see Fig.9),
those parameters present some dependence 
upon $\rho$ and $\eta$: full curve corresponds to $\rho = 0.09, \eta = 0.323$; while dashed one is
related to $\rho = 0.254, \eta = 0.442$. A visible discrepancy among these two curves is seen. 
\end{itemize}

\newpage

\section{Perspectives and conclusion}

$ \;\;\;\; \bullet $  Thanks to the HQET approach and the OPE formalism which is used, we have at our 
disposal rigorous and
{\it complete calculations} of the dynamics of the $B^{0 {\pm}}$ decays into two vector mesons. This 
formalism is available for all charmless $B$ decays provided the spin of the intermediate resonance(s) is
less or equal $1$; the only changes which must be taken into account are the $V_{CKM}$ matrix elements, the
masses and the widths of the resonances involved in each decay. 
\newline

$  \bullet $  In the case of leptonic decay of one resonance, like $J/{\Psi} \rightarrow e^+ e^{-}, {\mu}^+
{\mu}^-$, the angular distributions are modified because of the spin $1/2$ final leptons; which require
the use of other Wigner rotation matrices. Those calculations have been already done in our first paper
\cite{ZJA}. 
\newline

$ \bullet$ In the case where a $(c {\bar c})$ bound state or a charmed meson is produced like:

$$ B^0 \to  J/{\Psi} {\rho}^{0}, D^* X (X = {\rho}^{0}, \omega, K^{*0}), $$ 

the Wilson coefficients involved in the effective hamiltonian have to be modified, but we do not expect big
change with respect to the $c_i (c_i')$  coefficients used in the present paper. 
\newline

$  \bullet$ Other interesting consequences arise from this formalism: it can be easily extended to the numerous
channels like: $B \to  VP, \  PP \ $  where one or two pseudoscalar mesons ($P = 0^{-+}$) are produced
directly from the $B$ decay. Because of the simple equality
${\lambda}(P) = {\lambda}(V) = 0$, 
the number of helicity states is reduced from $3$ to $1$. 
\newline

$ \bullet$ An important point which has been mentionned in the present note is the role of the
 ${\rho}^0  -  {\omega}$ {\it mixing} and its consequence for the determination of the direct CPV parameter
 (Section  $3$ and reference \cite{AWTHOMAS}). Tagging of $B^+$ and $B^-$ is made easy thanks to the 
$K^+$ and $K^-$ mesons coming from the cascade decays. In our opinion, we can also exploit all the 
angular distributions of the final particles (and their correlations) in order to detect an 
eventual discrepancy which can arise between the $B^+$ and $B^-$ decays respectively. 
However, a complete study of those channels and their simulations require the knowledge of the strong phase
shift $\delta$ (mentionned in Section  3) according to the $\pi \pi$ invariant mass. Work is in progress. 
\newline

$  \bullet$ Those calculations and simulations can be implemented into {\bf SICBMC}, the Monte-Carlo generator
of the LHCb experiment, in order to perform afterwards a full analysis of the
simulated channels. 

\newpage

 \subsubsection*{Acknowledgements}
 
The authors are very grateful to Dr P.Perret, leader of the LHCb Clermont-Ferrand team, for his advices and
his suggestions.
\par 
One of us (Z.J.A.) is very indebted to Professor A.W.Thomas, 
Director of the Special Research Centre for the Subatomic Structure of Matter, for the very
exciting and illuminating discussions he got with him
about the QCD penguin diagrams and their importance in the evaluation
of the $B^0$ decay width.  
\vskip 0.5cm
This work was supported in part by the Australian Research Council and the University of Adelaide. 

\newpage

\section*{Appendix}

\appendix

%\subsection*{ 1) Polarizations in $B^{0}_{d}$ rest frame}

\section{Polarizations in $B^{0}_{d}$ rest frame}

%%%\chapter{Polarizations in $B^{0}_{d}$ rest frame}

$\;\;\;\;$ Momentum:
$$\vec{k}_{K} = -\vec{k}_{\rho} = \vec{k} = \left(
                               \begin{array}{c}
                                 k\sin\theta\cos\phi\\
                                 k\sin\theta\sin\phi\\
                                 k\cos\theta
                                \end{array}
                               \right), $$

where $\theta$ and $\phi$ are respectively polar and azimuthal angles of the produced $K^{*0}$.\\

Longitudinal polarization:\\ 
$$\epsilon_{K}(0) = \left(\frac{|\vec{k}|}{m_{K}},\frac{E_K}{m_K}\hat{k}\right), \ \ \ \epsilon_{\rho}(0) = \left(\frac{|\vec{k}|}{m_{\rho}},\frac{E_{\rho}}{m_{\rho}}(-\hat{k})\right).$$\\

Tranversal polarizations :
$$\vec{\epsilon}_{K}(1) = \left(
                               \begin{array}{c}
                                 \cos\theta\cos\phi\\ 
                                 \cos\theta\sin\phi\\ 
                                 -\sin\theta 
                                \end{array} 
                               \right) = \vec{\epsilon}_{\rho}(1),$$

$$\vec{\epsilon}_{K}(2) = \left(
                               \begin{array}{c} 
                                 -\sin\phi\\ 
                                 \cos\phi\\ 
                                  0 
                                \end{array} 
                               \right) = -\vec{\epsilon}_{\rho}(2).$$

Helicity frame :\\
$${\epsilon_{K}(+)}= \left(\epsilon(1) + i\epsilon(2)\right)/\sqrt{2}, \ \ \ 
{\epsilon_{K}(-)}= \left(\epsilon(1) - i\epsilon(2)\right)/\sqrt{2},$$\\

\vspace{-2.0em}

$$\vec{\epsilon}_{K}(+) = \left(
                               \begin{array}{c} 
                                 \cos\theta\cos\phi-i\sin\phi\\ 
                                 \cos\theta\sin\phi+i\cos\phi\\ 
                                  -\sin\theta 
                                \end{array} 
                               \right) / \sqrt{2} = \vec{\epsilon}_{K}^{\ *}(-) = \vec{\epsilon}_{\rho}(-),$$

$$\vec{\epsilon}_{K}(-) = \left(
                               \begin{array}{c} 
                                 \cos\theta\cos\phi+i\sin\phi\\ 
                                 \cos\theta\sin\phi-i\cos\phi\\ 
                                  -\sin\theta 
                                \end{array} 
                               \right) / \sqrt{2} = \vec{\epsilon}_{K}^{\ *}(+) = \vec{\epsilon}_{\rho}(+).$$

%%%\vskip 0.5 cm

%%%%%%%%%%%%%%%%%%%%%%%%%%%% ANNEXE 2 %%%%%%%%%%%%%%%%%%%%%%%%%%%%%%%

%\subsection*{ 2) Wilson's coefficients}
\section{Wilson's coefficients}

%%%\chapter{Wilson's coefficients}

We use, in the case of the $\rho^{0}$ production, the following linear combinations 
of the effective Wilson coefficients:
\begin{center}
\begin{eqnarray*}
& & \!  c_{t1}^{\rho}=c^{\prime}_{1} + \frac{c^{\prime}_{2}}{N_{c}}, \\
& & c_{p1}^{\rho}=-(c^{\prime}_{4} + \frac{c^{\prime}_{3}}{N_{c}})+\frac{1}{2}(c^{\prime}_{10}+\frac{c^{\prime}_9}{N_{c}}), \\
& & c_{p2}^{\rho}=\frac{3}{2}(c^{\prime}_7 + \frac{c^{\prime}_8}{N_{c}}+c^{\prime}_{9}+\frac{c^{\prime}_{10}}{N_{c}}),
\end{eqnarray*}
\end{center}
where $c_{t1}^{\rho}$ relative to tree diagram, $c_{pi}^{\rho}$ relative to penguin diagram and $0.98 <N_{c} <2.01$.\\

When $q^{2}/m^{2}_{b} = 0.3$:
\vspace{1em}
\newline
$ { \hspace{1.5em}}  c^{\prime}_{1} = -0.3125,  \ \ \  c^{\prime}_{2} = 1.1502,$     \\

$ c^{\prime}_{3} = 2.443 \times 10^{-2} + 1.543 \times 10^{-3}i, \ \ \  c^{\prime}_{4} = -5.808 \times 10^{-2} - 4.628 \times 10^{-3}i,$   \\

$  c^{\prime}_{5} = 1.733 \times 10^{-2} + 1.543 \times 10^{-3}i,  \ \ \  c^{\prime}_{6} = -6.668 \times 10^{-2} - 4.628 \times 10^{-3}i,$   \\

$  c^{\prime}_{7} = -1.435 \times 10^{-4} - 2.963 \times 10^{-5}i, \ \ \   c^{\prime}_{8} = 3.839 \times 10^{-4},  $ \\

$  c^{\prime}_{9} = -1.023 \times 10^{-2} - 2.963 \times 10^{-5}i,  \ \ \   c^{\prime}_{10} = 1.959 \times 10^{-3}.$   
\newline
\newline

When $q^2/m^2_b = 0.5$:
\vspace{1em}
\newline
$  { \hspace{1.5em}}  c^{\prime}_{1} = -0.3125, \ \ \  c^{\prime}_{2} = 1.1502,$ \\

$  c^{\prime}_{3} = 2.120 \times 10^{-2} + 2.174 \times 10^{-3}i,\ \ \  c^{\prime}_{4} = -4.869 \times 10^{-2} - 1.552 \times 10^{-2}i,$  \\

$  c^{\prime}_{5} = 1.420 \times 10^{-2} + 5.174 \times 10^{-3}i,\ \ \  c^{\prime}_{6} = -5.729 \times 10^{-2} - 1.552 \times 10^{-2}i,$ \\

$  c^{\prime}_{7} = -8.340 \times 10^{-5} - 9.938 \times 10^{-5}i,\ \ \  c^{\prime}_{8} = 3.839 \times 10^{-4},$ \\

$  c^{\prime}_{9} = -1.017 \times 10^{-2} - 9.938 \times 10^{-5}i,\ \ \  c^{\prime}_{10} = 1.959 \times 10^{-3}.$ 

%%%\vskip 0.5 cm

%%%\chapter{Form factors (BSW model)}

%\subsection*{3) Form factors (BSW model)} 

\section{Form factors (BSW model)} 
%%%%$$A_{i}^{B\to K^*,\rho}(m^2_{\rho,K^*})=\frac{h_i^{B\to K^*,\rho}}{1-\frac{m^2_{\rho,K^*}}{M_j^2}}$$

       \begin{center} 
            \begin{tabular}{|c|ccc|} 
               \hline[.3cm] 
               $ $ & $V$ & $A_1$ & $A_2$  \\ 
               \hline[.3cm] 
              $B\to K^* $ & $\frac{0.369}{1-{m^2_{\rho}(GeV^2)}/{5.43^2(GeV^2)}} $ & $\frac{0.328}{1-{m^2_{\rho}(GeV^2)}/{5.43^2}(GeV^2)} $ & $\frac{0.331}{1-{m^2_{\rho}(GeV^2)}/{5.43^2}(GeV^2)} $ \\ 
              \hline[.3cm]
              $B\to \rho $ & $\frac{0.329}{1-{m^2_{K^*}(GeV^2)}/{5.32^2}(GeV^2)} $ & $\frac{0.283}{1-{m^2_{K^*}(GeV^2)}/{5.32^2}(GeV^2)}$ & $\frac{0.283}{1-{m^2_{K^*}(GeV^2)}/{5.32^2}(GeV^2)} $ \\ 
               \hline[.3cm]
          \end{tabular} 
         \end{center}

  For further details, see reference \cite{AWTHOMAS} and literature quoted therein.

\newpage

%%%%%%%%%%%%%%%%%%%%%%%   BIBLIOGRAPHY  %%%%%%%%%%%%%%%%%%%%%%%%%%%%%%%%%%%%%%%

\end{document}